\documentclass[numberedappendix,iop]{emulateapj}

\usepackage{graphicx}
\usepackage{bm}
\usepackage{psfrag}
\usepackage{mathrsfs}

\begin{document}

\newcommand{\be}{\begin{eqnarray}}
\newcommand{\ee}{\end{eqnarray}}

%
%



\title{Assessing the Role of Spin Noise in the Precision Timing of Millisecond Pulsars}
\shorttitle{Timing Noise in Pulsars}

\author{
  Ryan~M.~Shannon and
  James~M.~Cordes}
  \shortauthors{Shannon \& Cordes}
\affil{Astronomy Department, Cornell University, Ithaca, NY 14853}
\email{ryans@astro.cornell.edu}
\email{cordes@astro.cornell.edu}


\begin{abstract}
We investigate rotational spin noise (referred to as timing noise) in non-accreting
pulsars: millisecond pulsars, canonical pulsars, and magnetars.  Particular attention
is placed on quantifying the strength and non-stationarity of timing noise in
millisecond pulsars because the long-term stability of these objects is required
to detect nanohertz gravitational radiation.  We show that a single scaling law
is sufficient to characterize timing noise in millisecond and canonical pulsars
while the same scaling law underestimates the levels of timing noise in magnetars.
The scaling law, along with a detailed study of the millisecond pulsar
B1937$+$21, leads us to conclude that timing noise is latent in most
millisecond pulsars and will be measurable in many objects when better
arrival time estimates are obtained over long data spans.   The sensitivity of
a pulsar timing array to gravitational radiation is strongly affected by any
timing noise.  We conclude that detection of proposed gravitational wave
backgrounds will require the analysis of more objects than previously
suggested  over data spans
that depend on the spectra of both the gravitational wave background and
of the timing noise.   It is imperative to find additional millisecond pulsars
in current and future surveys in order to reduce the effects of timing noise.

\end{abstract}


\section{Introduction}

In most pulsars the residual times of arrival (TOAs) 
show  structure that differs greatly from what is expected from measurement error alone
and is typically consistent with having a red power spectrum.  
This structure is generically referred to as spin noise or timing noise (TN).

Rotational irregularities of the neutron star appear to be the dominant source of TN in most pulsars.   
Timing noise is thought to arise from either changes in coupling between the neutron star crust and its superfluid core \cite[][]{1990MNRAS.246..364J} or
 magnetospheric torque fluctuations   \cite[][]{1987ApJ...321..805C,2006Sci...312..549K,2008ApJ...682.1152C,2010arXiv1006.5184L}.     
Thus the study of  TN provides valuable insight into the structure of the neutron star and its magnetosphere. 

The observed strength of  timing noise varies by more than eight orders of magnitude 
over the known non-accreting pulsars, which we subdivide into three classes: 
the  magnetars, with spin frequencies $\nu < 1/6$~s$^{-1}$ and relatively high magnetic fields; 
the rapidly-spinning and relatively weakly magnetized millisecond pulsars (MSPs), 
with spin frequencies $\nu > 50$~s$^{-1}$; 
and the canonical pulsars (CPs) with both spin frequencies and magnetic field strengths between the two other classes.  
 Some magnetars show root mean square  (rms) TOA variations of many tens of seconds on time scales of years, whereas the most stable MSPs have not shown evidence of TN at the $200$~ns level over decade-long time scales.

Millisecond pulsars, which have stability comparable to the best terrestrial clocks, continue to be intensely studied.   Their low levels of TN enable other TOA perturbations to be quantified, such as the relativistic effects in pulsars with massive (white dwarf) companions \cite[][]{2008ApJ...679..675V}, recoil from planetary-mass companions \cite[][]{2003ApJ...591L.147K},  and the presently-undetected stochastic background of gravitational waves \cite[][]{1979ApJ...234.1100D,1983ApJ...265L..39H,2006ApJ...653.1571J}.   
Interest in the detection of  gravitational waves with pulsars has intensified in recent years due to the improvement in MSP timing precision.
 This improvement can be attributed to technological advancements in telescope receivers and signal processing equipment \cite[][]{2007PhDT........14D}, improved analysis methods \cite[][]{2006ApJ...642.1004V}, and the discovery of new pulsars that appear to possess intrinsically superior timing stability \cite[][]{2006MNRAS.371..337O}.
   
      Long-term timing stability of millisecond pulsars is necessary to detect the small correlated perturbations in the TOAs associated with passage of gravitational waves through the solar system.
   It has been suggested that if sub$-100$~ns stability over $5-10$ years  can be achieved for a number of millisecond pulsars (currently estimated at $N_{\rm PTA}=20-40$) in a pulsar timing array (PTA),  
   a stochastic background of gravitational waves at a cosmologically significant level  can  be detected \cite[][]{2005ApJ...625L.123J}. 
Only two MSPs have shown any measurable TN, making characterizing as well as forecasting TN in MSPs difficult. 
However, the strength and properties of TN will certainly affect the detection of gravitational waves, even if TN is latent in most objects at present. 

In this paper we analyze TN throughout the pulsar population and assess the strength of TN in MSPs.
In  \S\ref{sec:TN_phen} we summarize the phenomenology of TN and show that random walk models and related non stationary processes can be used to model most observed TN.
In \S\ref{sec:TN_diags} we suggest two tools for diagnosing TN:  
   one  that is appropriate for assessing the long-term stability of MSPs and another that can be used to classify and compare TN throughout the pulsar population.      
In \S\ref{sec:TN}, 
we derive scaling relationships for TN in canonical pulsars, millisecond pulsars, and magnetars.
  We find that  millisecond pulsars have TN that is  consistent with that observed in canonical pulsars.
  We further link MSPs to CPs  by showing that the behavior of the MSP 
   B1937$+$21 is similar to that found in the canonical pulsar population. 
  In contrast, magnetars are found to possess TN that exceeds the amount expected from extrapolation from the other populations.
    In \S\ref{sec:GW}  we conclude that TN is present at levels that affect the observation strategies employed in pulsar timing arrays and suggest detecting gravitational radiation requires timing observations of more pulsars than previously estimated. 
 In that section we also discuss techniques for mitigating TN  and improving the sensitivity of a PTA to gravitational waves.

\section{Timing Noise:  Phenomenology}\label{sec:TN_phen}

Timing noise is manifested as structure in  the residuals of a fit to pulsar  TOAs. 
 A single TOA is determined by comparing a profile formed from averaging a large number of pulses with a template profile.  
The averaging is conducted to both increase the signal to noise ratio  
and decrease the effects of jitter associated with intrinsic pulse-to-pulse phase and amplitude variations.  
 The TOAs are then compared to a model that  accounts for the propagation of the pulse from the pulsar to the earth and refers the arrival times to the solar system barycenter \cite[][]{2006MNRAS.372.1549E}.  
 The fit includes terms accounting for periodic variations associated with the motion of the earth about the solar system barycenter and the reflex motion of the pulsar due to a companion, if the pulsar is in a binary system.   
The fit also includes secular terms that account for the unknown spin-down of the pulsar, and secular, but frequency-dependent terms that correct for the propagation of the radio pulses through plasma in the interstellar medium.  
It is essential to fit for the pulsar spin frequency $\nu$ and frequency derivative $\dot{\nu}$ because these quantities 
are intrinsic to the pulsar and cannot be predicted using any other technique. 
With the exception of a few young pulsars (such as the Crab and Vela pulsars) 
the values of the higher order frequency derivatives associated with pulsar braking are  
not measurable on the year to decade observing spans over which pulsars have presently been observed.


The residuals $\mathscr{R}(t)$ of the fit are used to assess the validity of the timing model and identify
the presence of unmodeled periodic and secular trends. 
The rms of the residuals   over an observing span of length $T$, after a second-order polynomial fit is given by
\be
\sigma_{\mathscr{R},2}^2 (T) = \frac{1}{N_t} \sum_i^{N_t} \mathscr{R}^2(t_i),
\ee  
 for an observation comprising $N_t$ samples at times $t_i$,  with $i=1,N_t$.

The variance $\sigma^2_{\mathscr{R},2}$  can be subdivided into a white component $\sigma_W^2$ 
and a red component  $\sigma_{{\rm TN},2}^2$ that in canonical pulsars is usually dominated by TN:
\be
\sigma^2_{\mathscr{R},2}(T) = \sigma_{{\rm TN},2}^2(T)  + \sigma_W^2.
\ee
In this discussion {\em red} is used to label processes which have ensemble average power spectra that have greater power at lower fluctuation frequencies (red spectra) and {\em white} for processes that have equal levels of power at all fluctuation frequencies (white or flat spectra).

There are a number of TN models that are distinguished by the time evolution of the residuals, or equivalently the shape of 
the power spectrum of the residuals.
 
One set of models is based on random walks of the spin properties of the pulsar.
 We consider random walks in pulse phase (${\rm RW}_0$), frequency (${\rm RW}_1$), and frequency derivative (${\rm RW}_2$) as useful archetypal processes. 
 Because the processes correspond to white noise in $\nu$, $\dot{\nu}$, and $\ddot{\nu}$,
they  produce rms residuals that scale proportional to $T^{1/2}$, $T^{3/2}$, and $T^{5/2}$, respectively \cite[][]{1972ApJ...175..217B}, 
and      ensemble average power spectra with spectral indices of $-2$, $-4$, and $-6$, respectively \cite[][]{1990ApJ...353..588H}.  
      For these processes the residuals have non-stationary statistics.  
      For reference, we note that a  gravitational wave background from merging massive black holes  produces rms residuals that scale proportional to $T^{5/3}$ and a power spectrum with a spectral index of ${-13/3}$ \cite[][]{2003ApJ...583..616J}.  
  
  Band-limited noise (BL) is associated with processes that have low and high frequency cut-offs, in which the rms residuals increase for some time with a slope dependent on the particular spectral shape of the process. 
 After a time associated with the low-frequency cut-off  of the band $T_{\rm out} = 1/f_{\rm low}$, the rms timing noise will plateau.  An example of a BL process is the perturbation induced by a wide asteroid belt around a pulsar (R.~Shannon et al., in preparation).

In many pulsars it appears that multiple types of TN occur at once. 
 However, random walks provide a good basis for modeling non-stationary components of timing noise.
 \cite{1985ApJS...59..343C} and \cite{1995MNRAS.277.1033D} conducted detailed analyses of complementary sets of canonical pulsars. 
While  they found that TN in most canonical pulsars cannot be explained by a single random walk process, both suggest that a mixture of random walks in $\nu$ and $\dot{\nu}$ and discrete jumps in $\phi$,  $\nu$, and $\dot{\nu}$ were compatible with the TN.   


Alternative models for timing noise include periodic and quasiperiodic processes. 
These models have gained favor because of recent reports of periodic and quasiperiodic contributions the the residual TOAs for a few pulsars.
For example, PSR B1931$+$24  \cite[][]{2006Sci...312..549K}  shows jumps between two states with distinct spin down rates $\dot{\nu}$ at quasiperiodic times.
 In a study of $366$ pulsars,  \cite{2010MNRAS.402.1027H} and \cite{2010arXiv1006.5184L} identify a few pulsars  ($\approx 2\%$ of their sample)  that contain  
periodic or quasiperiodic components in residual time series and switches between distinct states of $\dot{\nu}$. 
They also found that the different levels of $\dot{\nu}$ have unique average pulse profiles and they propose that this form of timing noise can be  corrected. 
In a substantial fraction of the identified cases of periodicity or quasiperiodicity, the model included a significant $\ddot{\nu}$ that is attributed to  non-stationary timing noise that augments any periodic or quasiperiodic component.
We  discuss the possibility  of mitigating timing noise further in \S\ref{sec:mit_TN}.

\begin{figure}[!ht]
\begin{center}
\begin{tabular}{c}
\includegraphics[scale=0.4]{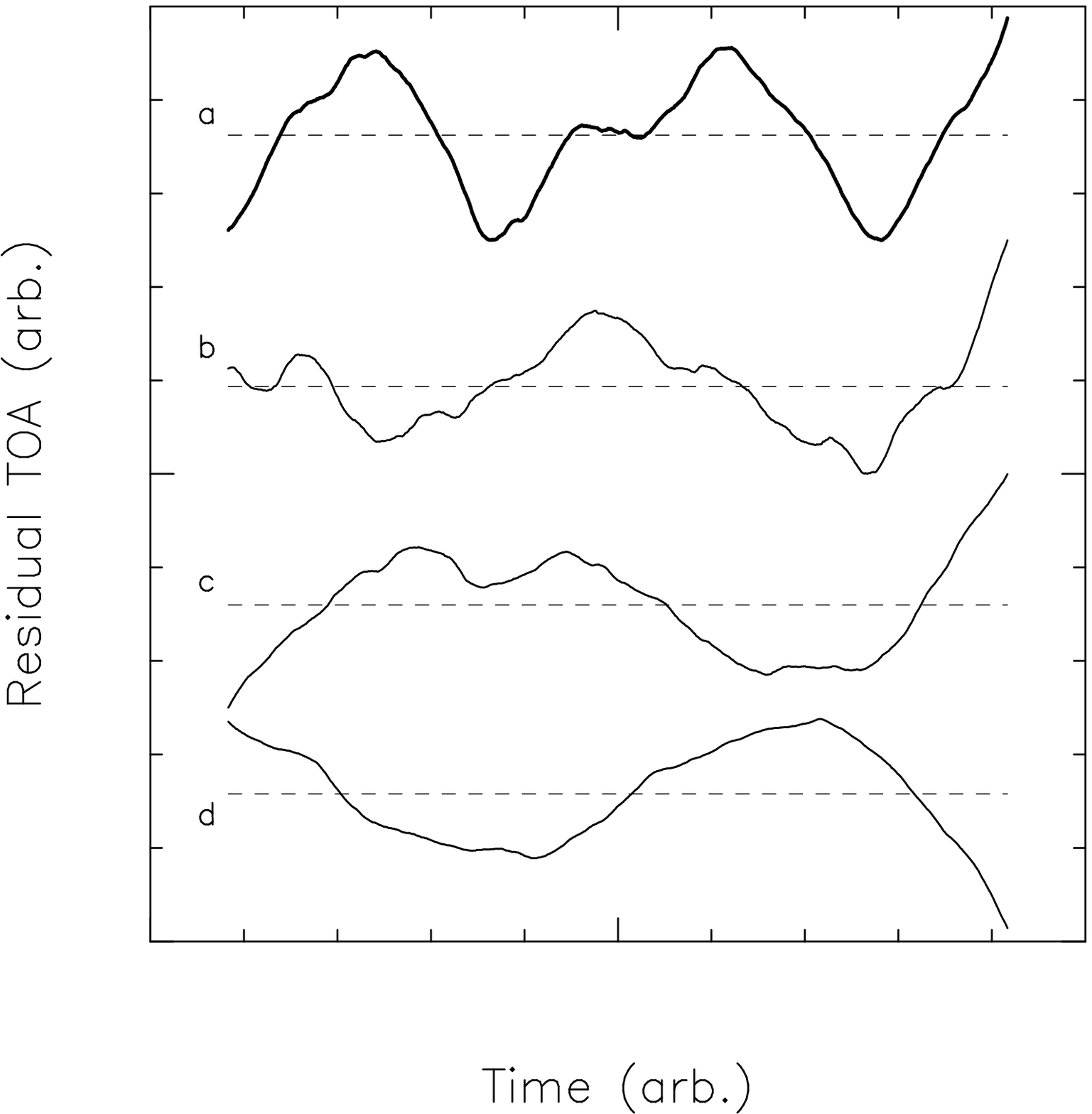}  \\
\includegraphics[scale=0.4]{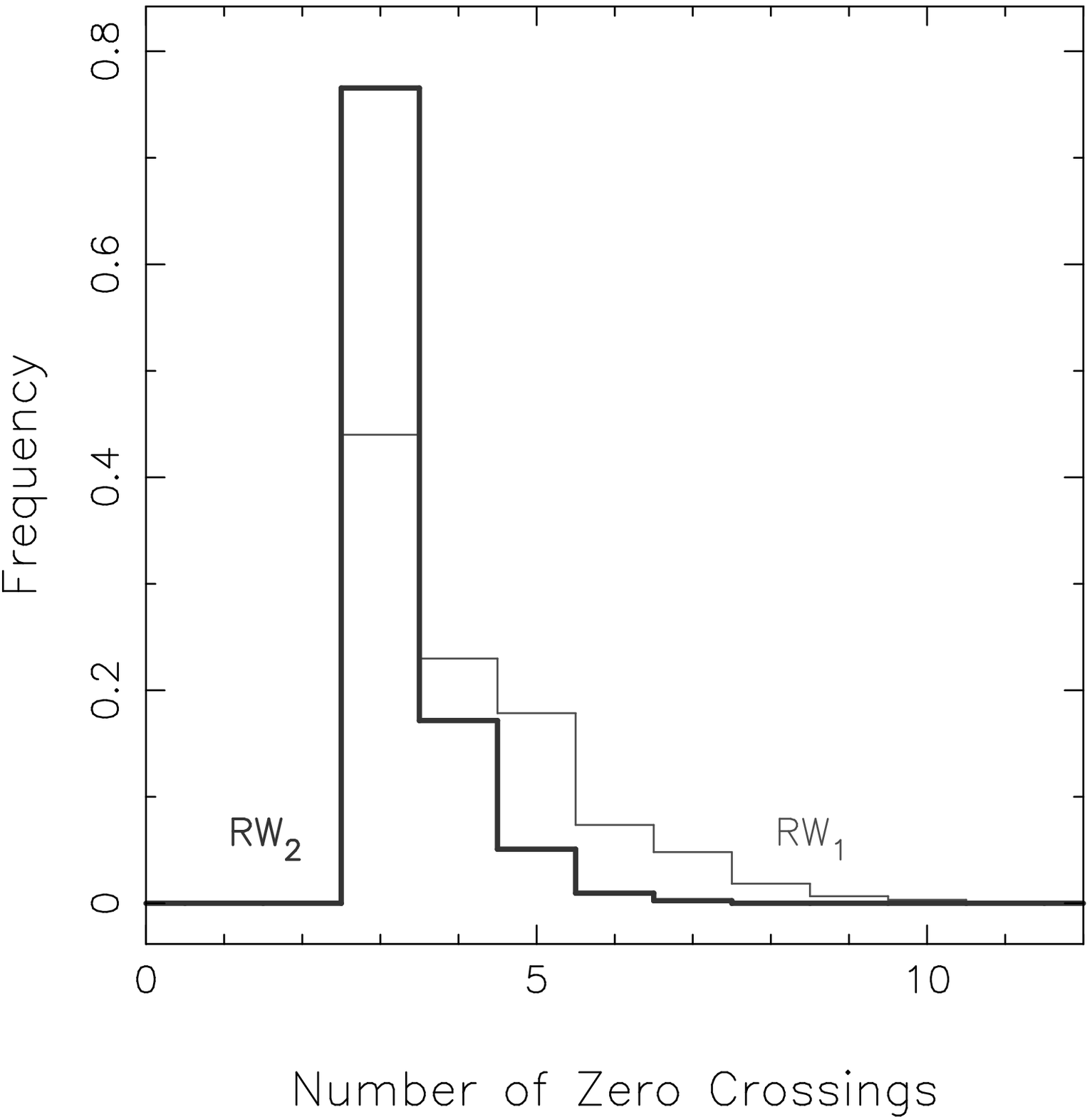} \\ 

\end{tabular}
\caption{ \label{fig:zero_crossing} {\em Upper panel:}  Four realizations of ${\rm RW}_1$ timing noise.  Curve {\em a} has a large number of zero crossings and behavior that could be misidentified as quasi-periodic in spectral analysis.   Curves {\em b} shows behavior that is irregular.  Curves {\em c} and {\em d} show behavior in which the cubic term is dominant, with $\ddot{\nu}>0$  in curve {\em c} and $\ddot{\nu} <0$  in curve {\em d}.   {\em Bottom panel:}  Histogram of the number of zero crossings for processes ${\rm RW}_1$ (thick lines) and ${\rm RW_2}$ (thin lines)  after including a quadratic fit.  Both  processes show realizations where the number of zero crossings is much larger than $3$ that mimic quasiperiodicity in spectral analyses.  The number of zero crossings for ${\rm RW}_0$ is off the scale of the graph. 
} 
\end{center}
\end{figure}

While in some cases, there is clear evidence of a periodic or quasiperiodic contribution to the TOAs, 
in other cases, realization to realization variation can mimic quasiperiodic behavior. 
To demonstrate this  we simulated residual curves for ${\rm RW}_0$, ${\rm RW}_1$, and ${\rm RW}_2$ random walks.  In the top panel of Figure \ref{fig:zero_crossing},  we show four realizations of quadratic-subtracted residual TOAs for the same ${\rm RW}_1$ process.
  In the plots residual curves show behavior that mimic quasiperiodicity, irregular behavior in which higher order polynomials dominate the TN, and cubic-dominated behavior with both $\ddot{\nu} >0$  and $\ddot{\nu}<0$. 
 
 We use the number of zero crossings to quantify the morphological variations in single realizations of RW processes.   
 Realizations that have a large number of zero crossings will appear quasiperiodic or irregular. 
 In contrast, realizations  that have three zero crossings will appear cubic and match what is expected from ensemble average statistics.  
 In the bottom panel of Figure \ref{fig:zero_crossing}  we show a histogram of the number of zero crossings for each quadratic-subtracted polynomial for $4000$ 	realizations of ${\rm RW}_1$, and ${\rm RW}_2$ processes.    
 A significant fraction of the realizations of both ${\rm RW}_1$ and ${\rm RW}_2$ processes show $>3$ zero crossings and a few show $> 6$ zero crossings.
 The number of zero crossings for residuals of ${\rm RW}_0$ random walks is not displayed, because 
 this random walk has  a shallow spectrum  and thus single realizations show very irregular behavior that typically have $> 10$ zero crossings.


\section{Timing Noise: Diagnostics}\label{sec:TN_diags}

Two approaches have been used to characterize the strength of timing noise in radio pulsars.   
The first uses the total  TN after a second order fit $\sigma_{{\rm TN},2}$. 
\cite{1980ApJ...239..640C} define the activity parameter as
\be
A = \log\left[\frac{ \sigma_{{\rm TN},2} (T)}{\sigma_{{\rm TN},2}(T)_{\rm Crab}} \right],
\ee
which measures levels of TN relative to the Crab pulsar and represents a time-independent measure of the strength of the TN, assuming that pulsars show TN
with similar time variability to the Crab pulsar.

A second set of methods characterizes timing noise using the frequency second derivative $\ddot{\nu}$ calculated from a cubic fit to the TOAs.
 Some groups have directly used $\ddot{\nu}$ to assess the strength of the TN \cite[][]{2006MNRAS.370L..76U, 2007ChJAA...7..521C} and correlated it with other pulsar parameters.
\cite{1994ApJ...422..671A}  assessed the strength of TN using a parameter 
\be
\Delta_8 = \log\left(\frac{|\ddot{\nu}|}{6 \nu} T_8^3\right),
\ee
where $\ddot{\nu}$ is measured over an observing span of $T_8 = 10^8$s.
While the cubic term will dominate the variance of TN in the ensemble average of any red process with a monotonically decaying spectrum,
 in a single realization higher order 
terms may contain a large portion of the TN. 
Thus statistics based on $\ddot{\nu}$ tend to underestimate the amount of TN in these processes.
Additionally, the statistic $\Delta_8$ is model-dependent because $\ddot{\nu}$ on average increases with length of observing span for red noise processes,  much like the total rms residuals.    
Therefore to properly compare values of $\Delta_8$ or $\ddot{\nu}$ in observations of different lengths a model-dependent time scaling needs to be included.

A dimensionless Allan variance-like parameter $\sigma_z$ is described in \cite{1997A&A...326..924M} that can be used to estimate pulsar stability, 
\be
\sigma_z(T) = \frac{1}{2\sqrt{5}}  \left[ \frac{\sigma_{\ddot{\nu}}(T)}{\nu} \right] T^2, 
\ee
where $\sigma_{\ddot{\nu}}(T)$ is the rms of $\ddot{\nu}$ over observing spans of length $T$. 
Because the parameter uses $\ddot{\nu}$ to estimate TN,  it also will in general underestimate the total TN.


Previous methods do not provide satisfactory diagnostics for TN.  
We therefore suggest that the rms timing noise (after a second order fit) is the basis for any proper diagnostic of TN, and 
propose two closely-related tools for diagnosing TN in pulsars.

To estimate the timing stability of a pulsar we use the post-fit rms TN scaled to $\nu$, $\dot{\nu}$, and time span $T$,
\be
\label{eqn:sigma}
\hat{\sigma}_{{\rm TN},2}  = C_2 \nu^\alpha |\dot{\nu}|^{\beta} T^\gamma,
\ee
where the parameters $C_2$, $\alpha$, $\beta$, and $\gamma$ are estimated
over the entire pulsar population.

A diagnostic  suitable for comparing timing noise across the pulsar population is the relative TN parameter 
\be
\label{eqn:zeta}
\zeta = \frac{\sigma_{{\rm TN},2}(T)}{\hat{\sigma}_{{\rm TN},2}(T)} = \frac{ \sigma_{{\rm TN},2}(T)}{C_2 \nu^{\alpha} |\dot{\nu}|^{\beta} T^{\gamma}},
\ee
which is the measured TN $\sigma_{{\rm TN},2}$, normalized by the global fit $\hat{\sigma}_{{\rm TN},2}$ from equation (\ref{eqn:sigma}).
The relative TN parameter is similar to the activity parameter $A$, 
but instead of normalizing to the properties of one pulsar (i.e., the Crab pulsar), the TN is compared to the best fit across all objects.   
This statistic can be used to identify outlying objects.  
  If a pulsar shows $\zeta \ll 1$ it has smaller levels of TN than expected.  
If  a pulsar shows $\zeta \gg 1$, it produces larger levels  of TN than expected. 
We note that because this parameter depends on the modeled timing noise it depends on the observations included in the fit.  
The parameter values will change when more objects are included in the fit, or objects are included over longer observing
spans.  As a result $\zeta$ will change when additional observation of  TN are included.  If the new observations have statistically similar behavior as the present observations of TN, the fit will not change in a significant way.  
If the additional objects have different behavior (for example, if timing noise became stationary over very large $T$), the revised values of $\zeta$ will better identify the outlying objects.

\section{Timing Noise across Neutron Star Populations}\label{sec:TN}


In this section, we show how rotational TN varies  across the canonical pulsar, millisecond pulsar and magnetar populations.  Previous analyses of TN have focused on canonical pulsars and fit for only a limited number of parameters using the statistical tools described in the previous section.    Instead we will use $\hat{\sigma}_{{\rm TN},2}$ and $\zeta$, which are defined in equations (\ref{eqn:sigma}) and (\ref{eqn:zeta}), respectively. 

\begin{figure*}[!ht]
\begin{center}
\begin{tabular}{cc}
\includegraphics[scale=0.5]{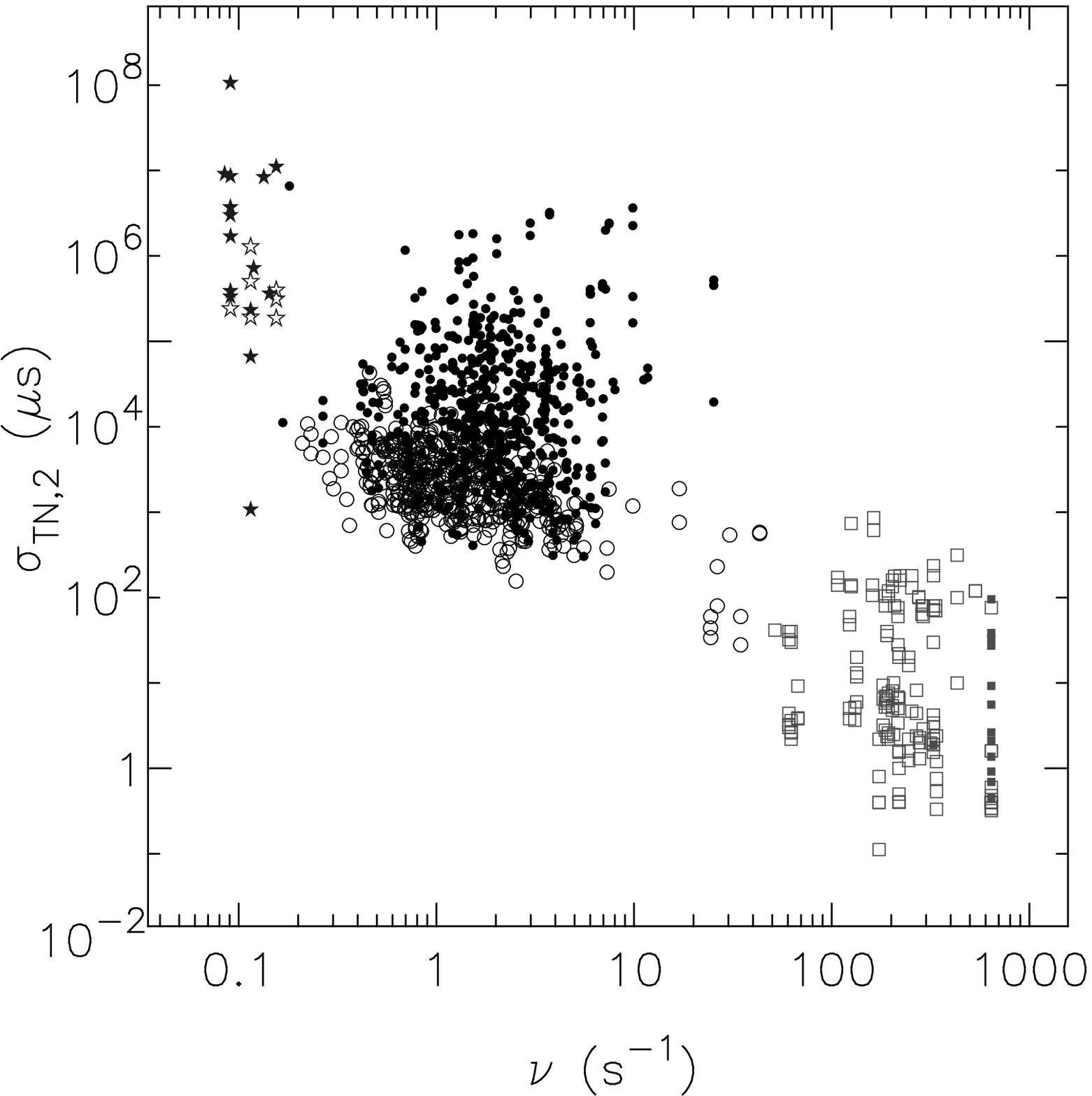} & 
 \includegraphics[scale=0.5]{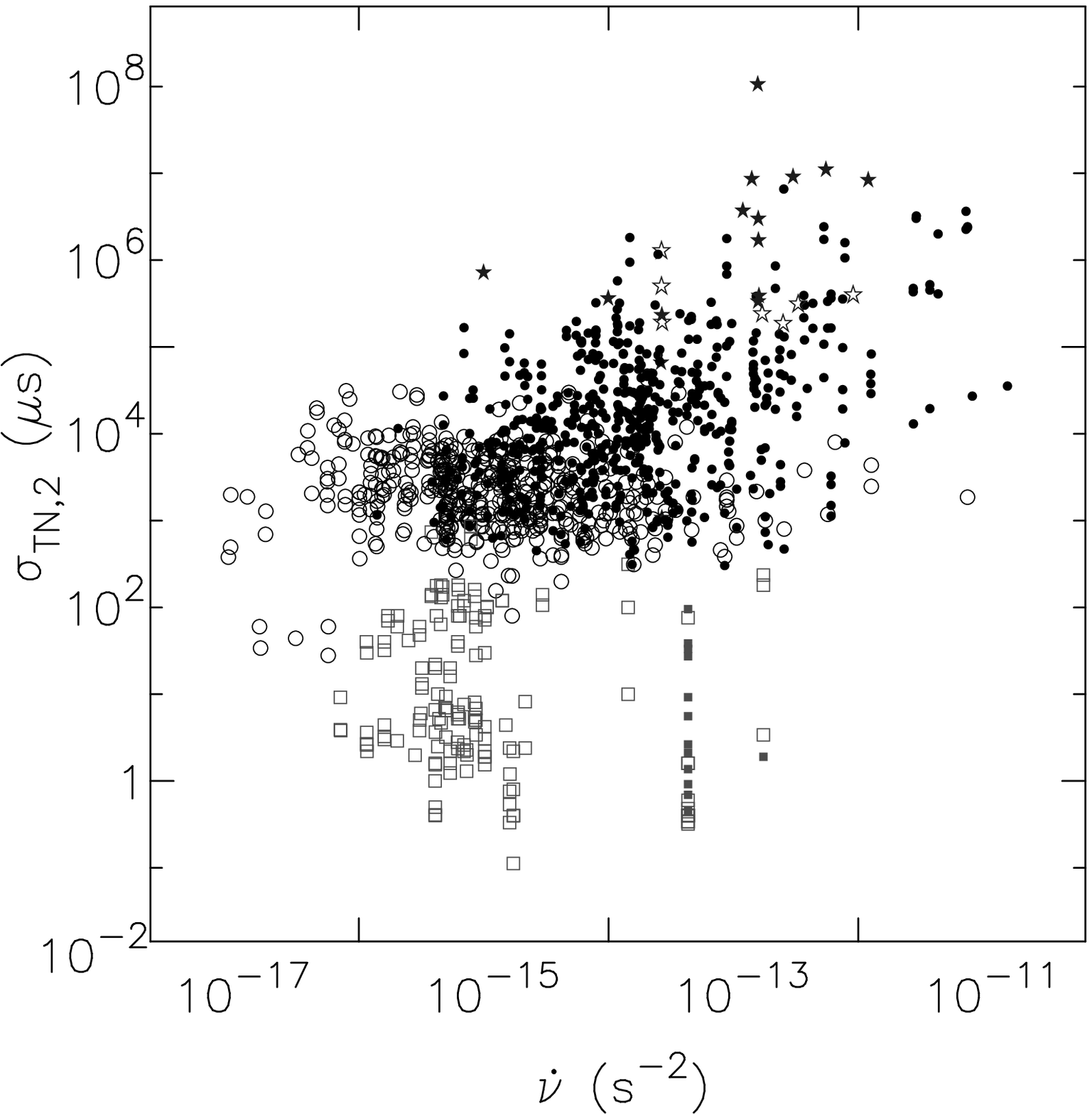} 
\end{tabular}
\caption{ \label{fig:sigma_corr} Scatter plots showing the correlation between measured rms timing noise $\sigma_{{\rm TN},2}$ with spin frequency $\nu$ (left panel) and frequency derivative, $\dot{\nu}$ (right panel).   Filled symbols represent detections of timing noise, and open symbols represent $2\sigma$ upper limits.  Magnetars ($\nu < 1/6$~${\rm s}^{-1}$) are identified by  stars, canonical pulsars ($50~{\rm s}^{-1} < \nu <1/6$~s$^{-1}$) are identified by  circles, and millisecond pulsars ($\nu >$~20~s$^{-1}$) are identified by  squares.  The observations encompass a wide range of observing spans $0.1~{\rm yr} < T < 30~{\rm yr}$.}
\end{center}
\end{figure*}

For our analysis we compiled observations of TN from many sources in the literature. 
In Appendix \ref{sec:campaign_list} we present the  observing campaigns that we use and describe our methods for  calculating $\sigma_{{\rm TN},2}$. 
Our analysis includes  $N_t= 1213$ time series, from approximately $450$~distinct pulsars, which include $N_D=591$ detections of TN and $N_{\rm UL}=622$ upper limits.  Our analysis excludes young objects that have measured frequency second derivatives $\ddot{\nu}$ that are attributed to pulsar braking. 
Plots displaying the rms timing noise $\sigma_{{\rm TN},2}$ versus $\nu$ and $\dot{\nu}$ are displayed in Figure \ref{fig:sigma_corr}. 

\subsection{Maximum Likelihood Analysis}\label{sec:meth_obs}


We use a maximum likelihood approach following \cite{1989tns..conf..119D} to find the best fit parameters for  equation (\ref{eqn:sigma}) in logarithmic space
\be
\label{eqn:log_model}
\ln(\hat{\sigma}_{{\rm TN},2}) = \ln{C_2} + \alpha \ln(\nu)+  \beta \ln | \dot{\nu}_{-15}|  + \gamma \ln(T_{\rm yr}), 
\ee
with  $\sigma_{{\rm TN},2}$ expressed in $\mu s$, $\nu$ expressed in s$^{-1}$, $\dot{\nu}_{-15}$ expressed in $10^{-15}~{\rm s}^{-2}$, and $T_{\rm yr}$ expressed in years.

  A fifth parameter $\delta$ is incorporated in the analysis to account  for the large scatter in the strength 
  of the timing noise.  
This scatter is associated with both realization to realization variation and non-modeled parameters 
that are assumed to be independent of $\nu$ and $\dot{\nu}$, such as neutron star mass and other physical elements of TN.

We assume that $\sigma_{{\rm TN},2}$ is log-normally distributed; therefore the probability density function (PDF) of measuring rms residuals $\sigma_{{\rm TN},2,i}$ is
\be
\label{eqn:TN_PDF}
f_{\sigma_{\rm TN}}(\sigma_{{\rm TN},2,i}) = \frac{1}{\sqrt{2 \pi \delta^2}} \exp \left[-\left(\frac{\ln(\hat{\sigma}_{{\rm TN},2,i}/\sigma_{{\rm TN},2,i})^2}{2 \delta^2} \right)  \right], 
\ee
where $\hat{\sigma}_{{\rm TN},2, i} = \hat{\sigma}_{{\rm TN},2, i}(C_2, \alpha, \beta, \gamma, \delta)$ is the modeled red noise component.  
  We define the probability $P_i$ as the product of the PDF and the measurement error
    \be
  P_i =  f_{\sigma_{\rm TN}}(\sigma_{{\rm TN},i}) \Delta(\ln \sigma_{{\rm TN}, i}) =   f_{\sigma_{\rm TN}}(\sigma_{{\rm TN},i}) \frac{\Delta \sigma_{{\rm TN}, i}}{\sigma_{{\rm TN},i}},
  \ee
 which assumes that the measurement error is small relative to $\delta$, a situation that is confirmed below.

We also incorporate upper limits from many observations using 
 the probability 
\be
P_{{\rm UL},i} = 1 - \frac{1}{2} {\rm erfc} \left[ \frac{\ln (\hat{\sigma}_{{ \rm TN} ,i}/\sigma_{{\rm TN},i})}{\delta \sqrt{2} } \right],  
\ee
where ${\rm erfc}$ is the complementary error function.
The total probability for $N_D$  detections of timing noise and $N_{\rm UL}$ upper limits is then
\be
\label{eqn:last_model_eqn}
P(C_2,\alpha, \beta, \gamma, \delta)  = \prod_i^{N_D}  P_i \prod_j^{N_{\rm UL}}  P_{{\rm UL},j}.
\ee

For each population, the probability space was examined using a series of grid searches.
An initial search was conducted with a  coarse grid and a wide range of values in each parameter 
 to identify the best-fit location and determine if multiple values of any of the parameters were allowed.
Refined grid searches were conducted with much narrower ranges in values  with fine gridding
 to calculate parameter estimation error and covariance.  

\subsection{Canonical Pulsars}\label{sec:cp}

A fit restricted to only the canonical pulsars yields well determined values of the parameters in equation~(\ref{eqn:sigma}).
We find significant correlation of the strength of timing noise 
with $\nu$, $\dot{\nu}$, and $T$. The estimated parameter values and their respective $\pm2\sigma$ ($95\%$)  confidence intervals  are presented in Table \ref{tab:parameters}.
The scaling of timing noise with observing span ($\gamma = 1.9 \pm 0.2$) is found to be intermediate to scalings expected from ${\rm RW}_1$ and ${\rm RW}_2$, for which we would expect $\gamma =3/2$, and $\gamma=5/2$, respectively.

Realization to realization variation associated with a stochastic process provides insufficient scatter account for the spread in timing noise that is characterized by the fit parameter $\delta = 1.6 \pm 0.1$.   
We simulated a large number of realizations of random walks ${\rm RW}_0$, ${\rm RW}_1$, and ${\rm RW}_2$  and determined that realization to realization variation will induce  a scatter in each process of $\delta=0.23$, $0.46$, and $0.60$, respectively.
We conclude that the inferred value of $\delta$  includes additional contributions from the actual TN processes that are not captured by single idealized random walk models.
 

These findings generally agree with previous studies of timing noise in canonical pulsars
that have concluded both that TN typically
shows non-stationary behavior characterized by a red power spectrum
and have established correlations between timing noise and other spin parameters.
\cite{1980ApJ...239..640C} found a correlation between the activity parameter $A$ and period derivative $\dot{P} = -\dot{\nu}/\nu^2$ in $50$ pulsars.  
\cite{1989tns..conf..119D} found a correlation between this activity parameter and $P$ and $\dot{P}$ in observations of $40$~canonical pulsars.

\begin{figure}[!ht]
\begin{center}
\includegraphics[angle=-90,scale=0.5]{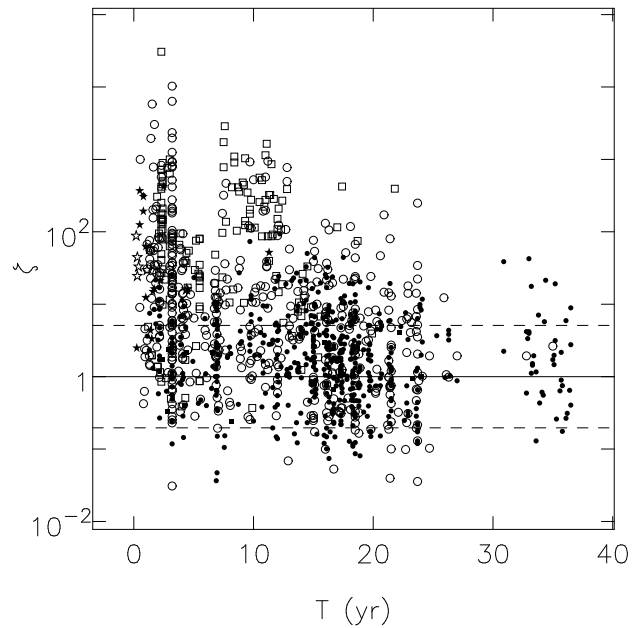} 
\caption{ \label{fig:res_tspan} Scatter plot showing the timing noise parameter  $\zeta$ versus observing span $T$.  We have used the timing noise model of the joint CP+MSP fit to calculate $\zeta$. There is no evidence for a change in timing noise characteristics over longer observing spans.  Filled symbols represent detections of timing noise, and open symbols represent $2\sigma$ upper limits.  Magnetars ($\nu < 1/6$~${\rm s}^{-1}$) are identified by  stars, canonical pulsars ($50~{\rm s}^{-1} < \nu <1/6$~s$^{-1}$) are identified by circles, and millisecond pulsars ($\nu >$~20~s$^{-1}$) are identified by  squares.    The solid lines indicates $\zeta =1$.  The dashed lines are the $\pm 1 \sigma$ variation of $\zeta$, as inferred from value of $\delta$ inferred from the joint CP+MSP fit. }     
\end{center}
\end{figure}

The scaling law models the timing noise over the entire range of observing spans and we find no evidence for band-limited  timing noise.
In Figure~\ref{fig:res_tspan} we display the relative TN parameter $\zeta$ versus observing span.    
If timing noise was band-limited over current observing spans, 
the amount of timing noise would plateau, and at large $T$,  the fit would be poor and  $\zeta \ll 1$.    In addition we found consistency between fits to CP observations with $T < 10$~yr and $T > 10$~yr. 


Analysis of a large on-going timing campaign 
of $366$ pulsars at the Jodrell Bank Observatory  with observing spans of  $10$ to $36$ years is presented in \cite{2010MNRAS.402.1027H}.
They calculated a scaling relation between $\sigma_z(10~{\rm yr})$ and $\nu$ and $\dot{\nu}$
\be
 \hat{\sigma}_z(10~{\rm yr}) = 10^{-11.5} \nu^{-0.4} |\dot{\nu}_{-15}|^{0.8},
\ee
where $\nu$ is measured in s$^{-1}$ and $\dot{\nu}_{-15} = 10^{-15}$~s$^{-2}$.
Our scaling relationship $\sigma_{{\rm TN},2} \propto \nu^{-0.9 \pm 0.2} |\dot{\nu}|^{1.0\pm 0.05}$ is inconsistent with this.  
In their analysis, $\sigma_z$ includes  contributions from additive white noise. 
If we conduct our analysis with $\sigma_{\mathscr{R}, 2}$ (i.e., include the white noise) instead of only to the red component $\sigma_{{\rm TN},2}$, we find a more consistent scaling relationship of $\sigma_{\mathscr{R},2} \propto \nu^{-0.7 \pm 0.1}|\dot{\nu}|^{0.76 \pm 0.02}$.


\subsection{Millisecond Pulsars}\label{sec:B1937}

\begin{figure*}[!ht]
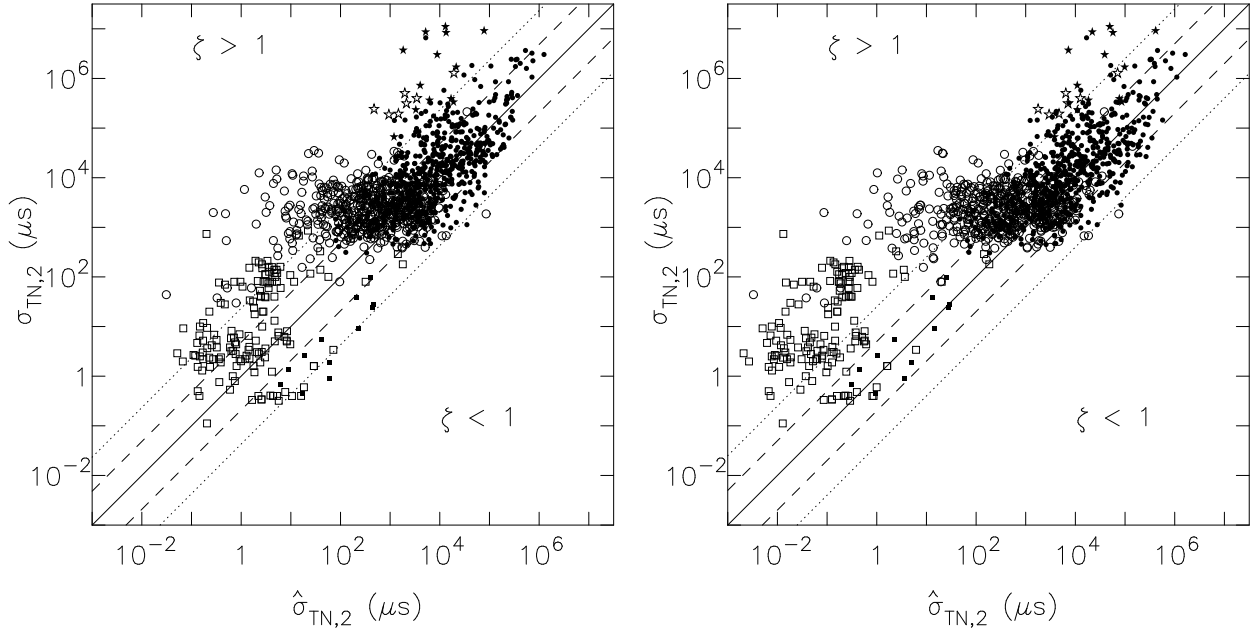

\begin{center}
\begin{tabular}{cc}
\includegraphics[angle=-90,scale=0.5]{f4a.eps} 
 &\includegraphics[angle=-90,scale=0.5]{f4b.eps}
\end{tabular}
\caption{ \label{fig:comp_rms} Correlation between predicted rms  TN $\hat{\sigma}_{{\rm TN},2}$ and measured rms TN $\sigma_{{\rm TN},2}$ for the CP-only model (left panel) and the joint CP+MSP model (right panel).   
Filled symbols represent detections of timing noise, and open symbols represent $2\sigma$ upper limits.  
 Magnetars ($\nu < 1/6~{\rm s}^{-1}$) are identified by stars, canonical pulsars ($50~{\rm s}^{-1} < \nu < 1/6$~s$^{-1}$) are identified by circles, and millisecond pulsars (MSPs) are identified by squares.
 Points above this line are observations that have levels of timing noise greater than expected ($\zeta > 1$).  
 Points below this line are observations that have  levels of TN less than expected by the model ($\zeta <1$).  The dashed lines show the expected width as estimated by the parameter $\hat{\delta}$, corresponding the $\pm1\sigma$ ($67\%$) width.  The dotted lines show $\pm2\sigma$ ($95\%$) width.   
 The CP-only model overestimates the strength of the timing noise in the MSPs, and both fits underestimate the level of timing noise in the magnetars. 
   }
\end{center}
\end{figure*}

Only two  MSPs have shown significant levels of timing noise: PSR B1937$+$21 (discussed in detail below), and PSR B1821$-$24. \cite[][]{2009MNRAS.400..951V}.  
 The model for canonical pulsars over-predicts the level of timing noise observed in PSRs B1937$+$21 and B1821$-$24, as displayed in Figure \ref{fig:comp_rms}.  
For both of these objects, the observed levels of TN are below the levels expected from the  
CP-only fit by a factor of one to two times $\delta$.

The best fit to the MSP population, listed in Table~\ref{tab:parameters}, has larger fitting uncertainties because the few observations of TN and constraining upper limits are restricted 
to smaller ranges in $\nu$, $\dot{\nu}$, and $T$.    
The fit is dominated by the many observations of timing noise in PSR~B1937$+$21.  
For a few MSPs, observations provide restrictive upper limits, but for many  the expected levels of timing noise are not constraining at the levels predicted by the CP-only fit.


We also conducted a joint fit of the MSP and CP populations.  In Figure \ref{fig:comp_rms} (right panel) we plot the observed levels of TN versus the levels predicted from the joint fit.
Visual inspection suggests 
that this fit provides a good model of the timing noise in the MSP population because the levels of timing noise observed are within the $\pm1\delta$ band and the upper limits exceed levels predicted by the model.   

The quality of fit is quantified by comparing observed and  predicted levels of timing noise using a $\chi^2$ statistic 
\be
\label{eqn:chi2_fit}
\hat{\chi}^2=  \sum_{i}^{N_D}  \frac{(\ln \sigma_{{\rm TN},2,i}-\ln \hat{\sigma}_{{\rm TN},2,i})^2}{\hat{\delta}^2},  
\ee
where only the $N_D$ observations of detected timing noise are included and upper limits are excluded.
If a model provides a good fit to the data, $\hat{\chi}^2$ follows a $\chi^2$ distribution.   For a fit to any population, the number of degrees of freedom is $N_{\rm DOF} = N_D-5$ if the population included in the fit (because $5$ parameters are included in the fit), and $N_{\rm DOF}=N_D$ if the population was not included in the fit.   
In Table~\ref{tab:fit_comp}, we list the values of $\hat{\chi}^2$ and corresponding probabilities $P$ that each fit models the individual populations.  This analysis confirms that the joint CP$+$MSP fit is a good model for both the CP and MSP populations.

  
The similarity of timing noise in MSPs to that in canonical pulsars is strengthened by examining the timing residuals of PSR B1937$+$21 in greater detail.
In terms of statistical precision, PSR B1937$+$21 is the best MSP in which to study timing noise because it shows the largest levels of TN of  any MSP. 

 In order to assess the strength and type of timing noise in PSR~B1937$+$21, we investigate how $\sigma_{{\rm TN},2}$  scales with observing span by combining the results  of many timing programs presented in Appendix \ref{sec:campaign_list}.      
In Figure \ref{fig:tn_1937}, the  rms residual timing noise is plotted versus observing span length for the various campaigns.    In this figure we also show model curves for random walks ${\rm RW}_0$, 
 ${\rm  RW}_1$, and ${\rm RW}_2$ scaled to an ensemble-average rms of $2~\mu$s over an $8$~year observing span, combined in quadrature with a $0.15~\mu$s white noise component, which matches the levels of noise in the short time span observations displayed in Figure \ref{fig:tn_1937}.  
  Over short time spans, the  residuals are dominated by white noise associated with instrumental sensitivity, 
pulse averaging effects, and diffractive interstellar scintillation \cite[][]{1990ApJ...349..245C}.
   



Inspection of this plot shows that the scaling of $\sigma_{{\rm TN},2}$ with $T$ is intermediate to ${\rm RW}_1$ and ${\rm RW}_2$ and therefore the scaling of TN with time is consistent with the observed scaling in the CP population ($\sigma_{\rm TN} \propto T^{2 \pm 0.2}$).
This scaling is inconsistent with ${\rm RW}_1$ or ${\rm RW}_2$ random walks. 
We note however that the power law scaling is altered if the amplitude of the RW steps have a power-law distribution \cite[for further discussion see Appendix C of ][]{1985ApJS...59..343C}.
The level of TN does not plateau over large $T$ so we conclude that  
the timing noise shows no sign of being band-limited on the current observation time scales.


%

Observational bias has lead to detection of TN in only two MSPs.   
The expected levels of timing noise in the other MSPs are below current observing sensitivity.   
Timing noise is observed in  PSR B1937$+$21 because it has a considerably larger $\dot{\nu}$ than other intensely studied MSPs. 






\begin{figure}[!ht]
\begin{center}
\includegraphics[angle=-90,scale=0.5]{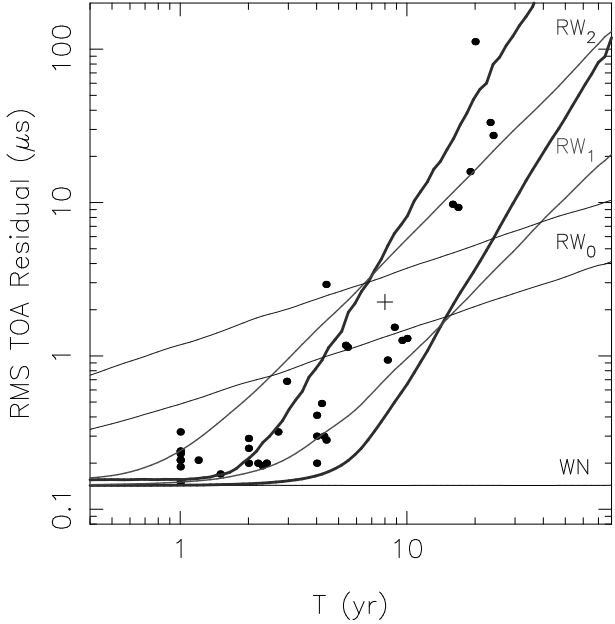} 
\caption{ \label{fig:tn_1937} Plot of the rms residuals $\sigma_{\mathscr{R},2} = \sqrt{\sigma_{{\rm TN},2}^2 + \sigma_W^2}$ versus observing span $T$ for PSR~B1937$+$21 and simulated random walks.  The large scatter in the observations at $T=1$~yr is associated with variable levels of white noise across timing programs.
The expected variation for random walks in phase $\phi$ (${\rm RW}_0$, thinnest lines), frequency $\nu$ (${\rm RW}_1$, medium thickness lines), and frequency derivative $\dot{\nu}$ (${\rm RW}_2$, thickest lines) are also displayed.  The $95\%$ confidence limits (based on simulations of a large number of realizations) are shown for each process.  The strength of the random walks are normalized to $\sigma_{{\rm TN},2} =2~\mu$s at $T=8$~yr, which is indicated by  the cross on the plot.  To each curve, white noise with rms strength of $\sigma_W = 0.15~\mu$s is added.  This level is denoted by the horizontal line marked ${\rm WN}$.  
 }
\end{center}
\end{figure}

\subsection{Magnetars}

The probabilities that the models fit the observed TN in the magnetar population are displayed in Table \ref{tab:fit_comp}.
  We find that the magnetar-only model provides a good fit to the observations (not surprisingly) but all other models under-predict the timing noise  in the magnetar population.
We conclude that magnetars show timing noise levels in excess of those found in the other populations of neutron stars.






\begin{deluxetable*}{lcccccl}
\tabletypesize{\footnotesize} \tablecolumns{7}
 \tablecaption{Best Fit Parameters\label{tab:parameters}}
\tablehead{ \colhead{Fit} & \colhead{$\ln({C_2})$} &    \colhead{$\alpha$} &  \colhead{$\beta$} &  \colhead{$\gamma$} &  \colhead{$\delta$}  & \colhead{$N_D (N_{\rm UL})$ } 
}
\startdata
CP&$2.0 \pm 0.4$  & $-0.9 \pm 0.2$ & $1.00 \pm 0.05$ & $1.9 \pm 0.2$ & $1.6 \pm 0.1$ &  563 (470) \\
MSP & $-20\pm 20$  &$1 \pm 2$ &  $2 \pm 1$ & $2.4 \pm 0.6$  & $1.2 \pm 0.5$ & 12 (147) \\
CP+MSP & $1.6 \pm 0.4$ & $-1.4 \pm 0.1$  & $1.1 \pm 0.1$ & $2.0 \pm 0.2$  & $1.6\pm 0.1$  &   575 (617)\\
MAG &   $3 \pm 7$ & $-1 \pm 3$  & $1.5 \pm 0.6$ & $3 \pm 1$ & $2.1 \pm 0.7$ &  15 (7) \\
CP+MAG &   $2.4\pm 0.5$ & $-1.4 \pm 0.2$  & $1.13 \pm 0.07$ & $ 1.7\pm 0.2$ & $1.7\pm 0.2$ &  578 (477)   \\
ALL  & $2.2 \pm 0.4$ & $-1.5\pm 0.1$ & $1.2 \pm 0.1$ & $1.8 \pm 0.1$ & $1.7 \pm 0.1$  & 590 (624) 
   \enddata
 \tablecomments{Best fit parameters and $\pm 2\sigma$ confidence limits for different populations of pulsars.   $N_D$ is the number of time series with detected timing noise used in the fit.  $N_{\rm UL}$ is the number of time series with upper limits of timing noise used in the fit.  
  }
  \end{deluxetable*}

\begin{deluxetable*}{crrrlrrlrrlrrlrrlrrl}
\tabletypesize{\footnotesize} \tablecolumns{20}
 \tablecaption{Fit Comparisons\label{tab:fit_comp}}
\tablehead{ \colhead{} & \colhead{} & \multicolumn{18}{c}{Model Fit} \\
\cline{3-20}\\
\colhead{} & \colhead{} & \colhead{} & \multicolumn{2}{c}{CP} & & \multicolumn{2}{c}{MSP} & &\multicolumn{2}{c}{MAG} & & \multicolumn{2}{c}{CP+MSP} & &\multicolumn{2}{c}{CP+MAG} & &\multicolumn{2}{c}{ALL} \\
\cline{4-5}  \cline{7-8}  \cline{10-11} \cline{13-14} \cline{16-17} \cline{19-20}  \\
\colhead{Family} & \colhead{$N_D$} & & \colhead{$\chi^2$} & \colhead{$P$}  & & \colhead{$\chi^2$} & \colhead{$P$} & & \colhead{$\chi^2$} & \colhead{$P$} & & \colhead{$\chi^2$} & \colhead{$P$} & & \colhead{$\chi^2$} & \colhead{$P$} & & \colhead{$\chi^2$} & \colhead{$P$}}
\startdata
CP & $563$ & & $532$ & ${0.8}$ & & $10^{5.2}$ & $(10^{-20})$ &  & $1996$  & $(10^{-20})$ & & ${538}$ & ${0.72}$ & & ${477}$ & ${0.99}$  & & ${493}$ & ${0.98}$  \\
MSP & $12$ & & $37$ &  $10^{-3.6}$ & & ${14}$ & ${0.05}$ & & $38$ & $10^{-3.8}$ & & ${3.8}$ &   ${0.8}$ & & $3.7$ & $0.99$ & &${7.0}$ & ${0.4}$ \\
MAG & $15$ & & $176$ & $(10^{-20})$ & & $10^{4.1}$ & $(10^{-20})$ & & ${10}$ & ${0.4}$ & &$98$ & $(10^{-10})$ & & $63$ & $10^{-7.1}$  & & $54$ &   $10^{-7.5}$   
 \enddata
  \tablecomments{Goodness of fit estimates for the canonical pulsars (CP), millisecond pulsars (MSP), and the magnetars (MAG) for models of sub-populations.  For the model fits the $N_D$ detected time series were used to calculate a $\chi^2$ statistic to assess the goodness of fit for the subgroups of the pulsar population.   Using this statistic, we calculated the probability $P$ that a fit modeled the observed levels of timing noise. Probabilities in parentheses are upper limits. 
  }
 \end{deluxetable*}

\subsection{Discussion: Timing Noise in Pulsar Populations}

There are  physical reasons to expect  timing noise in  MSPs and CPs to be consistent and follow a combined power law.
Magnetic fields almost certainly play a role in generating timing noise.  If surface or magnetospheric magnetic fields play the dominant role (likely the case in the case of a magnetospheric origin of timing noise, and plausibly the case if TN in associated with crust-core interactions), then timing noise would depend on pulsar spin parameters.   
Differences in the relationship between magnetic field strengths and the spin parameters $\nu$ and $\dot{\nu}$ between the populations would cause  a break-down in the scaling relations.
For example, if timing noise is associated solely with the neutron star core there may be a breakdown in the scaling relationship  because both CPs and MSPs may well have similar internal magnetic field strengths despite different spin characteristics.

Ultra-strong magnetic fields may cause the excess timing noise observed in magnetars. 
 Unlike canonical pulsars and millisecond pulsars, magnetar radiation appears to be driven by the decay of  magnetic fields, with some theories suggesting that the radiation is associated with crust cracking \cite[][]{2005Natur.434.1098H} that may be enhanced
compared to CPs. 
This cracking could drive rotational irregularities that contribute to the observed excess in timing noise.

Many radio pulsars have been discovered that have $\nu$ and $\dot{\nu}$ approaching those of the magnetars.
Additional timing observations of high magnetic field radio pulsars are needed to properly assess the difference between radio pulsars and magnetars.





\section{Implications for Gravitational Wave Detection} \label{sec:GW}

The presence of timing noise will significantly affect the sensitivity of a pulsar timing array 
to gravitational radiation.  At present, most MSPs show residuals consistent with  white noise.
Based on the scaling laws derived in \S\ref{sec:TN}, we predict that TN will be identified in many objects when they are monitored over longer time spans or observed with higher precision.

In Table \ref{tab:TN_PTA}, we list the MSPs that at present show the best timing stability.  
We show the expected rms timing noise over $2$~yr, $5$~yr, and $10$~yr observing spans, based on the scaling relationships discussed in  \S\ref{sec:TN}. 
  We also show the  $\pm 1\sigma$ variation that would be expected based on the observed spread of timing noise, and  measured limits on the amount of TN over $10$~yr observing spans.    
 For these pulsars TN is likely present at the $10$~ns to $100$~ns level  and 
 will therefore affect the detection of other TOA perturbations with amplitudes at these levels, such as a gravitational wave background. 
   In addition, we show predicted and measured levels of timing noise for  PSR B1937$+$21.  The large levels of timing noise imply this object will not contribute to the sensitivity of a PTA to GWs.   

 In the following we examine in detail the effect of the presence of timing noise on the properties of the PTA.

\begin{deluxetable*}{lrrrrrrrrrrrrrr}
\tabletypesize{\footnotesize} \tablecolumns{15}
 \tablecaption{Expected Levels Timing Noise for PTA Pulsars\label{tab:TN_PTA}}
\tablehead{ \colhead{} & \colhead{}  & \colhead{} &\multicolumn{3}{c}{$T = 2$~yr} &\colhead{} &\multicolumn{3}{c}{$T = 5$~yr} &\colhead{} &\multicolumn{4}{c}{$T = 10$~yr} \\
\cline{4-6}  \cline{8-10}  \cline{12-15} \\
\colhead{Object} & \colhead{$\nu$} &  \colhead{$\dot{\nu}$} & \colhead{$\hat{\sigma}_{\rm TN}$} &  \colhead{$\hat{\sigma}_{{\rm TN},L}$} &  \colhead{$\hat{\sigma}_{{\rm TN},U}$} & \colhead{} & \colhead{$\hat{\sigma}_{\rm TN}$} &  \colhead{$\hat{\sigma}_{{\rm TN},L}$} &  \colhead{$\hat{\sigma}_{{\rm TN},U}$} &  \colhead{} & \colhead{$\hat{\sigma}_{\rm TN}$} &  \colhead{$\hat{\sigma}_{{\rm TN},L}$} &  \colhead{$\hat{\sigma}_{{\rm TN},U}$}  &\colhead{$\sigma_{{\rm TN},{\rm meas}}$}\\
 \colhead{} & \colhead{(s$^{-1}$)} &  \colhead{($10^{-15}{\rm s}^{-2}$)}  
 & \colhead{(ns)} &  \colhead{(ns)} &  \colhead{(ns)} &\colhead{} 
 & \colhead{(ns)} &  \colhead{(ns)} &  \colhead{(ns)} & \colhead{} 
 & \colhead{(ns)} &  \colhead{(ns)} &  \colhead{(ns)} & \colhead{(ns)}}
\startdata
J0437$-$4715 & 174 & -1.73 &35 & 7 & 180 & & 210 & 41 & 1100 & & 830 & 160 & 4300 & $< 200$ \\ 
J1713$+$0747 & 219 & -0.41 &5 & 1 & 26 & & 31 & 6 & 160 & & 120 & 23 & 630 & $<200$\\ 
J1744$-$1134 & 245 & -0.54 &6 & 1 & 31 & & 36 & 7 & 190 & & 140 & 27 & 730  & $<620$ \\ 
J1909$-$3744 & 339 & -1.62 &13 & 2 & 68 & & 79 & 15 & 410 & & 310 & 60 & 1600 & $<170$ \\ 
\hline
B1937$+$21 & 623 & -43.30 &230 & 44 & 1200 & & 1400 & 270 & 7200 & & 5500 & 1100 & & $1500$
\enddata
  \tablecomments{Estimated strength of timing noise for selected PTA pulsars and PSR B1937$+$21  over $2$~yr, $5$~yr, and $10$~yr observing spans based on the best-fit  model to the canonical pulsars and the millisecond pulsars  (as defined in Table \ref{tab:parameters}).   For each  pulsar we list the spin frequency $\nu$ and spin frequency derivative $\dot{\nu}$.  For each observing span we show the expected values $\hat{\sigma}_{{\rm TN}}$ and the $1\sigma$ upper and lower limits: $\hat{\sigma}_{{\rm TN},L}$ and $\hat{\sigma}_{{\rm TN},U}$, respectively.  The limits are formally the quadrature sum of the parameter that quantifies the scatter of the distribution $\hat{\delta}$ and the  estimation error associated with the model, but are dominated by $\hat{\delta}$.   We also present the measured timing noise $\sigma_{{\rm TN}, {\rm meas}}$  (or upper limits) over $\approx 10$~yr observing span. }
 \end{deluxetable*}

\subsection{Timing Noise \& PTA Sensitivity } \label{sec:sens_TN}


To estimate the level of timing stability required to detect GWs, we calculate the GW detection signal to noise ratio (SNR) using a particular detection scheme.
We note the resulting conclusions are general and are relevant to all detection methods, including methods that are implemented using either frequentist or Bayesian approaches \cite[][]{2005ApJ...625L.123J, 2009MNRAS.395.1005V}. 

\subsubsection{Best Case: Gravitational Waves and Timing Noise Only}

In order to assess the best possible case,
we first consider TOAs that contain {\em only} perturbations associated with gravitational waves and timing noise.   
  For each pulsar $k$ at observation epoch $i$, the time of arrival perturbation (before any fit) $s_{ki}$ is altered by the correlated component of the GWB passing through the solar neighborhood $e_{ki}$, the uncorrelated component of the GWB outside of the solar neighborhood $p_{ki}$, and uncorrelated TN $r_{ki}$: 
\be
s_{ki} =  e_{ki} + p_{ki} + r_{ki}.
\ee
The perturbations $e$ and $p$ have the same rms strength. 
We define the SNR in the time series to be the ratio of the rms amplitudes (after a second order fit) of the correlated portion of the signal (i.e., $e_{ki}$)  to the the uncorrelated portion of the signal ($p_{ki}+r_{ki}$).
Thus in the residuals from a single pulsar the SNR is at most unity and is smaller if TN is present.

We now consider one approach to GW detection that involves forming a coherent sum (R.~Shannon \& J.~Cordes, in preparation) of the 
residuals for $N_{\rm PTA}$ pulsars, which increases the SNR. 
The best case configuration is when all the pulsars are located in a small patch of the sky,
but at different distances away from the observer.  In this case $e_{ki}$ is completely correlated between pulsars  and $p_{ki}$ and $e_{ki}$ are uncorrelated.  
As a result,  $e_{ki}$ is amplified  relative to $p_{ki}$ and $r_{ki}$ 
by a factor $\sqrt{N_{\rm PTA}}$.   
The combined SNR in a single data block of span $T$ of observations from $N_{\rm PTA}$ pulsars is
\be
\label{eqn:SNR_AMPTS}
\left(\frac{S}{N} \right)_{T,1}  && = \sqrt{  \frac{ N_{\rm PTA} \sigma_{{\rm GW},2}^2(T)}{ \sigma_{{\rm TN},2}^2(T) + \sigma_{{\rm GW},2}^2(T)}},
\ee
where the rms strengths of the GWB and the TN are characterized by $\sigma_{{\rm GW},2}(T)$ and $\sigma_{{\rm TN},2}(T)$, respectively. 

A test statistic based on the coherent sum  has an SNR of 
\be
\label{eqn:SNR_TS}
\left(\frac{S}{N} \right)_{{\rm TS},M}  =  \sqrt{\frac{M N_{\rm PTA}} {1+ \sigma_{{\rm TN},2}^2(T_M) / \sigma_{{\rm GW},2}^2(T_M)}},  
\ee
where we have assumed the data set can be subdivided and 
 $M$ independent estimates of the TS can be calculated (for example by using data blocks of length $T_M = T/M$), resulting in an enhancement of the SNR by a factor of $\sqrt{M}$.
We note that there are alternative ways to subdivide the time series. 
 \cite{2006ApJ...653.1571J} decompose the residuals using a set of orthonormal polynomials and calculate a TS using each polynomial, while  \cite{2009MNRAS.400..951V}  conduct an analysis in the Fourier transform domain.  In all cases the optimal value of $M$ is limited by other sources of noise (like white noise), which we discuss further in \S\ref{sec:effect_WN}.


The scaling relationship of equation (\ref{eqn:SNR_TS}) is used to establish the properties of a PTA sufficient to detect the GWB.   
To detect the gravitational wave background with a strength $\sigma_{{\rm GW},2}(T)$ with ${\rm SNR}_{{\rm TS}, M} > S_{\rm min}$  requires that the TN in an individual pulsar satisfy
\be
\label{eqn:stab_req}
\sigma_{{\rm TN},2}(T_M) <  \sigma_{{\rm GW},2}(T_M)\sqrt{ \frac{M N_{\rm PTA}}{S_{\rm min}^2} -1}. 
\ee

The number of pulsars required to  detect a GWB of strength ${\sigma_{{\rm GW},2}}(T)$ with an SNR greater than $S_{\rm min}$  with TN at a level $ \sigma_{{\rm TN},2}(T)$  is
\be
\label{eqn:NPSR_required}
N_{\rm PTA} > \frac{S_{\rm min}^2}{M}  \left[1 + \left(\frac{\sigma_{{\rm TN},2}(T_M)}{\sigma_{{\rm GW},2}(T_M)} \right)^2 \right].
\ee

Here we make two preliminary estimates of the requirements for GW detection using equations (\ref{eqn:stab_req}) and (\ref{eqn:NPSR_required}).    In \S\ref{sec:num_psr} we give a more detailed assessment that uses the model of TN in the pulsar population presented in \S\ref{sec:TN}.

As a first example, we estimate pulsar stability requirements to detect the expected stochastic background of merging massive black hole (MBH) binaries.  
Stochastic GWBs are typically characterized by their expected strain response $h_c(f)$ and not $\sigma_{{\rm GW},2}$.  In Appendix \ref{sec:strength_GWB} we show how to calculate $\sigma_{{\rm GW,2}}$ from $h_c(f)$.
 The MBH  background is presently considered the strongest plausible GWB, and is expected to induce a strain response of $h_c(f) = A_0 (f/1~{\rm yr}^{-1})^{-2/3}$, where the value of $A_0$ is estimated to be between $10^{-16}$ and $10^{-15}$ \cite[][]{2003ApJ...583..616J,2010arXiv1001.3161S}. 
Over a $T=5$~yr observing span, the MBH GWB will contribute $\sigma_{{\rm GW},2} = 19~{\rm ns}~(A_0/10^{-15})$ to the times of arrival, as indicated in Table \ref{tab:NP_PTA_2}, which presents results for this section and for \S\ref{sec:num_psr}.    
To achieve a signal to noise ratio in the detection statistic of $S_{\rm min}=5$ for a PTA comprising $N_{\rm PTA}=40$~pulsars using $M=1$ observation blocks, 
 timing noise must be limited to $\sigma_{{\rm TN},2} (T=5~{\rm yr}) < (\sqrt{3/5}) \sigma_{{\rm GW},2}(T=5~{\rm yr}) \approx 15$~ns. 
 
 We can also estimate the number of pulsars required to detect a GWB if TN levels are equal to the amplitude $\sigma_{{\rm TN},2} \approx 20$~ns over $5$ years exhibited by the pulsars in Table \ref{tab:TN_PTA}. 
A PTA comprising $N_{\rm PTA}= 70$ pulsars would yield an SNR of $S_{\rm min}= 5$, assuming $M = 1$, and a GWB with the same properties as in the previous example.  
However, a number of MSPs are expected to have TN at levels below the scaling law and therefore the required number of pulsars may be somewhat lower, which is described in \S\ref{sec:num_psr}

\subsubsection{Effect of White Noise on the Number of Independent Sub-blocks}\label{sec:effect_WN}

In general, the number of subdivisions $M$ that maximizes the SNR of the TS depends on the amplitude of other  noise contributions to the residuals, in particular white noise, which is guaranteed to be present in pulsar timing observations.  
We define a time scale $T_{ M}$ over which the expected GW signal exceeds the white noise levels $\sigma_{W, {\rm TS}}$ in the coherent time series by the same threshold as the TN, i.e., $\sigma_{{\rm GW},2} (T_{M}) = S_{\rm min } \sigma_{W,{\rm TS}}(T_{M})$.    
 For a total observing span of length $T$ there are $M \approx T/T_{\rm M}$ independent data blocks if $T > T_{M}$; if not, there is $M=1$ data block.  

The random noise in  the TS associated with the WN is given by   
\be
\sigma_{W,{\rm TS}}(T) = \frac{\sigma_n}{\sqrt{N_{\rm PTA} N_{\rm obs}(T)}}, 
\ee
where $\sigma_n$ is the level of white noise in a single observation and $N_{\rm obs}(T)=R_{\rm obs} T$ is the number of observation epochs in the interval of length $T$, and is characterized by an observation rate  $R_{\rm obs}$ (or equivalently an observing cadence of $R_{\rm obs}^{-1}$).
For a background with $\sigma_{{\rm GW},2}(T)= \sigma_g T^{5/3}$, the minimum time is 
\be
T_M = \left( \frac{ \sigma_n}{ \sigma_g} \frac{ S_{\rm min}}{\sqrt{N_{\rm PTA} R_{\rm obs}}}\right)^{6/13}. 
\ee 

For an array of $40$ pulsars, with $N_{\rm obs}(T) = 10~T_{\rm yr}$ (i.e., 10 observations per year), $\sigma_n=100$~ns rms error  per residual, and a GWB with strain spectrum $h_c(f) = 10^{-15} (f/1~{\rm yr}^{-1})^{-2/3}$,  the minimum block size is $T_{\rm min} \approx 2$~yr for $N_{\rm PTA} =40$~to~$100$.  The minimum block length $T_{\rm min}$ is approximately the same for both values of $N_{\rm PTA}$ because of the weak dependence of $M$ on $N_{\rm PTA}$, $M \propto N_{\rm PTA}^{3/13}$, for the assumed GWB background.



\subsection{The Fraction of MSPs Suitable for PTAs}\label{sec:num_psr}

Using the stability requirements defined in equation (\ref{eqn:stab_req}), the fraction of MSPs suitable for inclusion in a PTA, $\mathscr{F}_{\rm MSP}$,  can be evaluated.
   This fraction  is equivalent to the probability that a pulsar within the population has rms timing noise less than some threshold amount $\sigma_t$  over an observing span of length $T$.  Based on our TN model, this probability is
\be \label{eqn:TN_frac_main_text}
&&P(\ln \sigma < \ln \sigma_t|T) = \int_{ -\infty \  }^{\ln \sigma_t}  d\ln \sigma \int d {\bm M} \rho_{\bm M}({\bm M}) \nonumber \\
&&~~\times \int d \nu d\dot{\nu} \rho_{\nu, \dot{\nu}}(\nu, \dot{\nu}) \rho_{\ln \sigma}  (\ln \sigma | {\bm M}, \nu,\dot{\nu}, T), 
\ee
where $\rho_{\bm M}({\bm M})$ is the PDF of the parameter distribution, $\rho_{\nu, \dot{\nu}}$ is the PDF of the pulsar distribution in $\nu$ and $\dot{\nu}$, and $\rho_{\ln \sigma} $ is the PDF of the level of timing noise, given the model parameters.

In Appendix \ref{sec:PTA_pulsars}, we present methods for evaluating equation (\ref{eqn:TN_frac_main_text}).
The sensitivities of PTAs comprising $N_p=40$ and $100$ pulsars with a variety of observing spans $T$ are investigated with the same conditions as in \S\ref{sec:sens_TN}.  We have modeled timing noise $\sigma_{{\rm TN},2}$ using the joint CP and MSP model as presented in Table \ref{tab:parameters} and equation (\ref{eqn:sigma}).


In Table \ref{tab:NP_PTA_2} we show the fraction of pulsars suitable for inclusion in these PTAs.  
The fraction of suitable pulsars is smaller at $T=5$~yr than $T=2$~yr because the level of expected timing noise has increased relative to the GW signal but the number of sub-blocks $M$ has not changed. 
To produce a PTA comprising $N_{\rm PTA}$ pulsars requires the investigation of a total sample of MSPs, $N_{\rm MSP} = N_{\rm PTA}/\mathscr{F}_{\rm MSP}$ pulsars.  
To calculate $M$ we have again assumed that the pulsars are observed $10$ times per year with an rms precision on a single TOA of $100$~ns.    
For a PTA comprising $40$ to $100$ high quality MSPs and data spans of $5$ to $10$ years, our analysis indicates that  a total MSP sample that is two to three times  larger than $N_{\rm PTA}$ needs to be investigated in order to identify high quality objects.





\begin{deluxetable}{rrrrrrrrr}
\tabletypesize{\footnotesize} \tablecolumns{9} 
\tablecaption{Timing Noise Constraints on MSPs Suitable for a PTA\label{tab:NP_PTA_2}}
\tablehead{ 
\colhead{}  & \colhead{} &  \multicolumn{3}{c}{ $N_{\rm PTA} = 40$ }  & \colhead{} & \multicolumn{3}{c}{$N_{\rm PTA} = 100$}\\
\cline{3-5}  \cline{7-9} \\
\colhead{$T$ } & \colhead{$\sigma_{{\rm GW},2}$} & \colhead{$M$} & \colhead{$\sigma_{{\rm TN},2,t}$} &   \colhead{$\mathscr{F}_{\rm MSP}$} & \colhead{} & \colhead{$M$} & \colhead{$\sigma_{{\rm TN},2,t}$} &   \colhead{$\mathscr{F}_{\rm MSP}$} \\
\colhead{(yr)} &  \colhead{(ns)} &  \colhead{} & \colhead{(ns)} & \colhead{($\%$)} &  \colhead{} & \colhead{} & \colhead{(ns)} & \colhead{($\%$)}  }
\startdata
$2$ & $ 4 $ &  $ 1 $ &  $ 3 $ &$30\pm7$  & &$ 1 $ & $ 7 $ & $46\pm8$ \\ 
$5$ & $ 19 $ &  $ 1 $ &  $ 14 $ &$25\pm6$  & &$ 1 $ & $ 32 $ & $40\pm7$ \\ 
$10$ & $ 59 $ &  $ 2 $ &  $ 28 $ &$37\pm7$  & &$ 3 $ & $ 31 $ & $56\pm7$ \\ 
$20$ & $ 187 $ &  $ 5 $ &  $ 34 $ &$50\pm7$  & &$ 6 $ & $ 45 $ & $64\pm7$\\
\hline
$5$ & $ 19 $ &  $ 2 $ &  $ 9 $ &$41\pm7$  & &$ 2 $ & $ 15 $ & $53\pm7$ \\
$5$ & $ 19 $ &  $ 4 $ &  $ 4 $ &$55\pm8$  & &$ 4 $ & $ 7 $ & $65\pm7$ 
\enddata
\tablecomments{Fraction of pulsars suitable for inclusion in a PTA, $\mathscr{F}_{\rm MSP}$, for arrays comprising $N_{\rm PTA} = 40$ and $100$ pulsars, over observing spans ranging from $T=2$~yr to $T=20$~yr.  We have assumed a background with characteristic strain spectrum of $h_c(f) = 10^{-15} (f/1~{\rm yr}^{-1})^{-2/3}$. We have also assumed a detection signal to noise ratio of $S_{\rm min}=5$, using equation (\ref{eqn:stab_req}), with $M$ independent data blocks,as described in the main text.
   Also listed are the quadratic-corrected rms contribution of the gravitational wave background $\sigma_{{\rm GW},2}(T)$ for the total observing span $T$  and the threshold TN level $\sigma_{{\rm TN},2,t}(T_M)$ for the sub-block length $T_M= T/M$.  
}
\end{deluxetable}

\subsection{Mitigating Timing Noise}\label{sec:mit_TN}

\begin{figure}
\begin{center}
\includegraphics[scale=0.5]{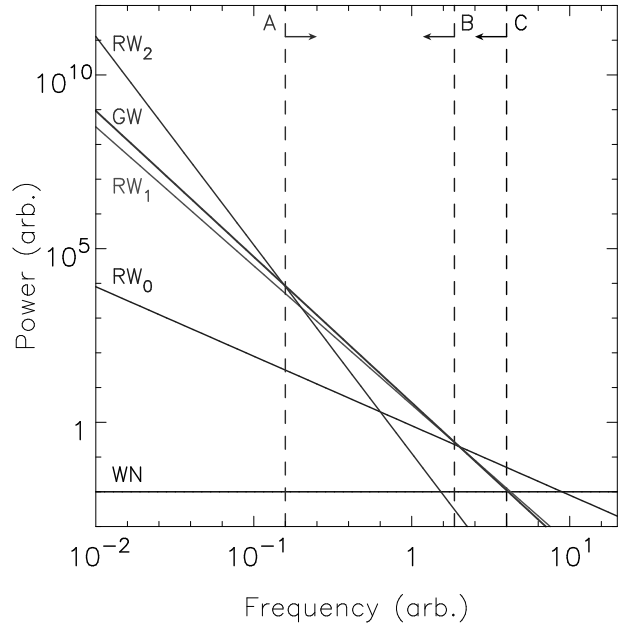} 
\caption{ 
\label{fig:spec_rms} 
Schematic power spectra for stochastic processes that contribute to pulsar timing residuals.  The dashed lines indicate fluctuation frequencies at which the TN exceeds the GWB for various processes and the arrows associated with these lines identify region in which the GWB signal is accessible.  Vertical dashed line {\em A} identifies the region in which the spin frequency  noise (${\rm RW}_2$) exceeds the gravitational wave background. Vertical dashed line {\em B} identifies the region in which phase noise (${\rm RW}_0$) exceeds the gravitational wave background.  Vertical dashed line {\em C} identifies the region in which white noise (${\rm WN}$) exceeds the gravitational wave background.       
} 
\end{center}
\end{figure}

It has been previously noted that applying a low-pass spectral filter to the residual TOA time series can improve the signal-to-noise ratio in a PTA \cite[][]{2005ApJ...625L.123J}. 
The presence and diversity of red noise in the MSPs  necessitates filtering that is tailored to the properties of each individual pulsar. 
Schematic  power spectra for ${\rm  RW}_0$, ${\rm RW}_1$, and ${\rm RW}_2$ random walks  and the gravitational wave background are displayed in Figure~\ref{fig:spec_rms}.    
 For systems in which ${\rm RW}_2$  is the dominant form of TN high-pass filtering (i.e, removing the lowest frequency components  of the signal) can be used to mitigate the contribution of TN to the TOAs.
It   is not possible to develop a  filter that mitigates ${\rm RW}_1$ timing noise without also removing the gravitational wave signal because they have very similar spectral shapes.

 There is evidence that pulse profile information can be used to correct residual time series. 
\cite{2010arXiv1006.5184L} identify a link between changes in pulse shape (probably connected to  mode changing) and changes in $\dot{\nu}$, and demonstrate that some timing noise can be corrected by identifying the time of the mode changes and the estimated values of $\dot{\nu}$.
Even if all TN is associated with mode changing  it is necessary to observe with a short cadence  to accurately determine the time at which the change occurred.     
\cite{2010arXiv1006.5184L} argue that daily observations would be necessary to adequately mitigate this mode changing timing noise in their objects.
If objects are not observed with a sufficiently short cadence, the uncertainty in the time of mode change introduces residual timing noise in the corrected time series with an amplitude proportional to the uncorrected level of timing noise. 
As noted in \S\ref{sec:TN_diags}, in half of the cases presented in \cite{2010arXiv1006.5184L}, the correlation was found after a fit for $\ddot{\nu}$ that removes non-stationary TN could not be corrected in their data.

\subsection{Future Prospects}\label{sec:GW_prospects}


A large number of pulsars need to be studied because timing noise will limit the utility of many objects and MSP timing stability cannot be fully constrained with spin parameters $\nu$ and $\dot{\nu}$.
As a result, if  $N_{\rm PTA}$ pulsars are required for a significant detection of the GWB,  a much larger number of pulsars ($N_{\rm MSP} = N_{\rm PTA}/\mathscr{F}_{\rm MSP}$, as outlined in \S\ref{sec:num_psr}) need to be discovered and characterized.   
To constrain the timing stability it is necessary  to conduct timing observations with sufficient precision to detect the presence of TN at the threshold level, as set by equation (\ref{eqn:stab_req}).  For realistic PTA configurations and a reasonable detection SNR, the required stability level will be at most a factor of a few greater than the anticipated strength of the gravitational wave background.  
An object is suitable for inclusion into the PTA if timing noise is not detected at this level.

Pulsar timing arrays can be expanded by both incorporating presently known objects with good intrinsic stability that are currently excluded due to low flux and discovering new MSPs with suitable timing stability.


Additional MSPs suitable for incorporation into a PTA are continually being discovered, with ongoing surveys with the Arecibo, Green Bank, Effelsberg, Parkes Telescopes; targeted searches for radio pulsar companions to {\em Fermi} gamma-ray point sources; and in the near future with the LOFAR Array \cite[][]{2010A&A...509A...7V}.
While occasionally bright MSPs have been discovered \cite[][]{2003ApJ...599L..99J}, selection effects generally  bias new discoveries toward fainter pulsars, and thus suitable objects require longer observations with more sensitive telescopes to  mitigate white radiometer noise. 

The requirements for finding and timing ultra faint MSPs highlight the need to use high-gain telescopes 
such as the Arecibo telescope and the proposed Square Kilometre Array (SKA). 
The SKA is estimated to find up to $\approx 6000$~MSPs \cite[][]{2009A&A...493.1161S}. 
If we conservatively estimate that $10\%$ of the MSPs are suitable, there will be $\approx 600$ objects available for inclusion in a PTA.  
The very best of these could comprise a PTA sufficient to {\em detect} the GWB while 
a much larger PTA could be used to study in detail the GWB and detect and examine individual GW sources.
Large interferometers such as the SKA  will be particularly important for improving throughput of timing campaigns because  they can be divided into sub-arrays that can observe multiple objects simultaneously.




\section{Conclusions}

We have developed scaling relationships for timing noise in millisecond pulsars, canonical
pulsars, and magnetars.   
We find that timing noise in MSPs is consistent 
with that observed in canonical pulsars. 
The timing behavior of the millisecond 
pulsar B1937$+$21 supports universality of TN in CPs and MSPs. 
Latent timing noise is predicted to be present in other MSPs with similar properties (but smaller) magnitudes 
to that in the CPs and PSR B1937$+$21, 
in accord with their smaller spin down rates.  
This timing noise may be measurable in
many pulsars when either longer data sets or higher precision arrival times are obtained.
Timing noise in magnetars is greater than that expected from extrapolation from the canonical pulsars.




\acknowledgements
We thank George Hobbs for providing a preprint of  \cite{2010MNRAS.402.1027H} in advance of publication 
and the referee for comments that improved the clarity of the text.
This work made use of NASA's ADS System   and the ATNF Pulsar Catalogue \cite[][]{2005AJ....129.1993M}.
This work was supported by the NSF through grant AST-0807151 and by NAIC, which is operated by Cornell University under a cooperative agreement with the NSF.

\vspace{0.1 in}

\appendix

\section{Reduction Procedure \& Timing Campaigns Used}\label{sec:campaign_list}

We have synthesized the results of many timing observations  to construct the scaling relationships for timing noise (TN) in canonical pulsars (CPs), millisecond pulsars (MSPs), and magnetars  (MAGs), as described in \S\ref{sec:TN}; and to conduct the case study of the MSP B1937$+$21 (as described in \S\ref{sec:B1937}).  
In Table \ref{tab:time_sources} we summarize the timing campaigns used in the analyses, the average length of data span contained in the campaign, and the number and type of objects analyzed.       

In the following sections we outline the procedure used to properly combine the results from all of the campaigns. 
In \S\ref{sec:ap_scaling} we describe how the root mean square (rms) timing noise $\sigma_{{\rm TN},2}$ is calculated from other TN diagnostics. 
In \S\ref{sec:ap_UL} we justify the threshold used for detecting the presence of red noise in timing data.
In \S\ref{sec:exclude}, we justify the exclusion of some timing observations from this study.
In \S\ref{sec:ap_1937}, we describe the observations used that form the basis of our study of PSR B1937$+$21.


\subsection{Calculating $\sigma_{{\rm TN},2}$}\label{sec:ap_scaling}

We use the  rms timing noise after a second order polynomial fit $\sigma_{{\rm TN},2}(T)$ as the primary observable property of TN, as justified in \S\ref{sec:TN_diags}.  

In Table \ref{tab:time_sources}, for each reference we identify the type of observations reported.  
While some timing campaigns report $\sigma_{{\rm TN}, 2}$ 
(coded ${\rm TN}$ in Table \ref{tab:time_sources}),  others report different but related measurements of TN.  
There are two notable conversions that are occasionally needed. 
Some campaigns report only the total rms residuals and the white noise level, and the red timing noise must be extracted from these quantities. For other campaigns, the timing noise is modeled in a functional form.   

{\em Calculating $\sigma_{{\rm TN},2}$ from the total timing noise and the white noise (Code {\em TW} in Table \ref{tab:time_sources}):}  Many timing campaigns report the total rms residuals and the levels of white noise in the observations.  In this case, the amount of timing noise is the quadrature difference between the rms residuals $\sigma_{\mathscr{R},2}$ and the white noise in the time series:
\be
\sigma_{{\rm TN}, 2}^2 =  \sigma_{\mathscr{R},2}^2 - \sigma_W^2.
\ee
In these observations the the level of white noise $\sigma_W$  reported comes from one of two sources: either the rms of residuals  after the TN has been analytically modeled;  or an estimate from the white noise in a single TOA.    

{\em Including modeled timing noise, (Codes S, STW, or H in Table \ref{tab:time_sources}):}
 In some cases the fit includes terms $\mathcal{M}_{\rm TN}(t)$  that model the timing noise.     The model is typically a series of polynomials or sinusoids.  It is typically included to provide an estimate of $\ddot{\nu}$ (in which case $\mathcal{M}(t) =  \ddot{\nu}(t-T_e)^3/6$, where $T_e$ is the epoch at which the spin properties are defined) or to improve the determination of modeled parameters of interest such as astrometric terms.  In these cases $\sigma_{{\rm TN}, 2}$ is approximated as the quadrature sum of the rms of $\mathcal{M}_{\rm TN}(t)$ and the TN contained in the post-fit residuals:
\be
\sigma_{{\rm TN},2}^2 =   \sigma_{{\rm TN},\mathcal{M}}^2(T)  + \frac{1}{T}\int_{t_0}^{t_1} \mathcal{M}_{\rm TN}^2(t) dt,  
\ee
where $t_0$ and $t_1=t_0 +T$ are the starting and ending epochs of the observations.  

For observations labeled {\em S}, polynomials have been used to model the timing noise.
For observations labeled {\em H}, harmonically related sinusoids have been used to model the timing noise. 
For observations labeled {\em STW}, a fit including $\ddot{\nu}$ was completed that partially whitens the residuals.  In addition to $\ddot{\nu}$, the  total rms timing noise $\sigma_{\rm TOT,\mathcal{M}}$ and the whitened rms timing noise $\sigma_W$ were reported.      
In this case, the rms timing noise is  $ \sigma_{{\rm TN},\mathcal{M}}^2(T) = \sigma_{\rm TOT,\mathcal{M}}^2 - \sigma_W^2$.


\subsection{The Detection Threshold for $\sigma_{{\rm TN},2}$}\label{sec:ap_UL}

 To determine if timing noise is detected in a time series, we conservatively require that the rms timing noise exceed twice the white noise floor (i.e.,  $\sigma_{{\rm TN},2}> 2\sigma_W$), because we suspect that in many timing programs the residuals were not examined at sufficient detail to rule out TN below this level.  
 If the measured TN in a time series does not meet the threshold we declare the time series to be an upper limit with a value of $2 \sigma_W$.        
 This is much larger than the formal detection threshold, $\sigma_{{\rm TN},2} > \sigma_{W}/\sqrt{N_{\rm DOF}}$, where $N_{\rm DOF}$ is the number of degrees of freedom in the residual TOAs.

\subsection{Excluded Observations}\label{sec:exclude}

We excluded observations of globular cluster pulsars which show acceleration (and  significant $\ddot{\nu}$) associated with these dense environments.
We have also excluded some additional reports of timing noise from this analysis:\\
{\em PSR J1012$+$5307.} \cite{2001MNRAS.326..274L} report a non-zero $\ddot{\nu}$ that they attribute to TN.  However a more recent analysis by \cite{2009MNRAS.400..805L} that includes the previous data shows no evidence for $\ddot{\nu} \neq 0$.  We therefore omit the measurement of \cite{2001MNRAS.326..274L}.\\
{\em PSR J1713$+$0747.}  \cite{2004PhDT........21S} report a non-zero $\ddot{\nu}$ that they attribute to TN.     However, a more recent analysis by \cite{2009MNRAS.400..951V} shows no evidence for $\ddot{\nu}\neq 0$.  We therefore exclude the measurement of \cite{2004PhDT........21S}. \\
{\em PSR B1937$+$21.}  We have also excluded a measurement of TN for PSR B1937$+$21 from this analysis, which is discussed in the next section.

\subsection{PSR B1937$+$21}\label{sec:ap_1937}

The observations used for the analysis of the scaling of the rms TN for PSR B1937$+$21 (discussed in \S\ref{sec:B1937})
 are presented in Table \ref{tab:tn_1937}.   We report the observing span of the observations the rms residuals $\sigma_{\mathscr{R},2}$, and, when available, the number of TOAs used in the analysis.  
In order to increase the number of independent observations at short observing spans, the publicly available residual TOAs from \cite{1994ApJ...428..713K} were subdivided into shorter observing spans of $1$, $2$, and $4$ years.
We note that many of the observations contain contemporaneous or common observations, and therefore many of the data points are not formally independent.

All the campaigns included  have been corrected for dispersion measure (DM) variations, determined by measuring the arrival time difference contemporaneously in two frequency bands.   
The DM correction is more accurate in recent campaigns because of improved observation procedures.  
In early campaigns, the measurements at two frequencies were performed many days apart and changes in interstellar propagation over those times likely increase TOA uncertainty.  In more recent campaigns, two-frequency observations often occur consecutively during the same observing session or simultaneously with dual-frequency receivers.

We have excluded the $12.5$~yr measurements of timing noise in PSR B1937$+$21 reported in \cite{2009MNRAS.400..951V} because the time series contained a long gap between observations with two different instruments. 
An arbitrary time offset between the two instruments (i.e., a jump) was included the fit.  
This jump removes a significant amount of TN from the residual time series.


\section{Estimating the fraction of suitable pulsars}\label{sec:PTA_pulsars}

In this section, we describe the methods for calculating the fraction of pulsars suitable for inclusion in the pulsar timing array, $\mathscr{F}_{\rm MSP}$. 
The fraction of pulsars that  show TN below a threshold RMS $\sigma_{{\rm TN},t}$ is equivalent to the probability of finding a pulsar within the population with those properties,
\be
\label{eqn:TN_frac}
P(\ln \sigma < \ln \sigma_t|T) = \int_{ -\infty \  }^{\ln \sigma_t}  d\ln \sigma \int d {\bm M} \rho_{\bm M}({\bm M}) \int d \nu d\dot{\nu} \rho_{\nu, \dot{\nu}}(\nu, \dot{\nu}) \rho_{\ln \sigma}  (\ln \sigma | {\bm M}, \nu,\dot{\nu}, T),
\ee
where $\rho_{\bm M}$ is the probability density for observing fit parameters, where ${\bm M} = (C_1, \alpha, \beta, \gamma, \delta)$, as in equation (\ref{eqn:log_model}); $\rho_{\nu, \dot{\nu}}$ is the probability density for the pulsar spin distribution; and $\rho_{\ln \sigma}$ is the PDF for observing a value of TN, assuming fixed values for the fit parameters.

We will assume that the level of TN $\sigma$ is log-normally distributed about the expected value:
\be
&&\rho_{\ln \sigma}  (\ln \sigma | C_1, \alpha, \beta, \gamma, \delta, \nu,\dot{\nu}, T)  = 
\frac{1}{\sqrt{2 \pi \delta^2}} \exp \left[ -\frac{(\ln \sigma/\hat\sigma)^2}{2 \delta^2} \right].
\ee
This is consistent with the analysis of \S\ref{sec:TN}, and the large observed spread in TN for pulsars with similar spin parameters $\nu$ and $\dot{\nu}$.

To model the $\rho_{\nu, \dot{\nu}}$, we will use the  observed distribution of MSPs:
\be
\label{eqn:pdf_spin}
\rho_{\nu,\dot{\nu}}(\nu, \dot{\nu}) =\frac{1}{N_p} \sum_p^{N_p} \delta(\nu - \nu_p) \delta(\dot{\nu} -\dot{\nu}_p).
\ee
For the analysis presented here, we use the $64$ non-globular cluster MSPs listed in the ATNF pulsar catalogue \cite[][]{2005AJ....129.1993M}.

The parameter space PDF $\rho_{\bm M}$ is modeled using estimates of the best fit values and the fitting covariance matrix ${\bm C}$
\be
\label{eqn:model_pdf}
\rho_{\bm M}({\bm M} | \ln C_1, \alpha, \beta, \gamma, \delta_T )  = \frac{1}{\sqrt{(2\pi)^{5} \det({\bm C}^{-1})}} \exp \left[ ({\bm M}-\hat{\bm M})^T {\bm C}^{-1}    ({\bm M}-\hat{\bm M})\right],
\ee
where $M = (\ln C_1, \alpha, \beta, \gamma, \delta)^T$  and $\hat{M}$ is a vector containing the best fit parameters to the joint CP+MSP fit.
For ease of computation, the PDF was approximated  using a large number $N_s = 10^5$ of parameter values drawn from equation (\ref{eqn:model_pdf}):
\be
\label{eqn:pdf_params}
&& \rho_{\bm M} = \frac{1}{N_s} \sum_s^{N_s} \delta(\ln C_1 -\ln C_{1,s}) \delta(\alpha -\alpha_s) \delta(\beta -\beta_s) \delta(\gamma-\gamma_s)\delta(\delta-\delta_s).  
\ee
To calculate $P$, equations (\ref{eqn:pdf_spin}) and (\ref{eqn:pdf_params}) were substituted into equation (\ref{eqn:TN_frac}).

To calculate the estimation error in $P$ associated with the fitting error in model ${\bm M}$, we analyzed the distribution of $P_i$ using single realizations of the parameters  to calculate $\rho_{\bm M}$, i.e., we substituted 
\be
\rho_{{\bm M},i} =    \delta(\ln C_1 -\ln C_{1,i}) \delta(\alpha -\alpha_i) \delta(\beta -\beta_i) \delta(\gamma-\gamma_i)\delta(\delta-\delta_i)
\ee 
into equation (\ref{eqn:TN_frac}) to calculate a number of realizations of the probability $P_i$.
The standard deviation of  $P_i$  is the estimation error.


\section{Strength of the Gravitational Wave Background} \label{sec:strength_GWB}

In this section,   we calculate the rms strength of the gravitational wave background in the residuals $\sigma_{{\rm GW},2}(T)$ for a strain amplitude $h_c(f)$. 
The former  quantity is the strength of the GW signal accessible to pulsar timing observations and is used to estimate the sensitivity of a PTA in \S\ref{sec:GW}.

The strain amplitude is usually modeled with power-law behavior over the range of $f$ relevant to pulsar timing observations,    
\be
\label{eqn:ap_strain_amp}
h_c(f) = A_0 \left( \frac{f}{f_0}\right)^\alpha,
\ee
and is characterized by an amplitude $A_0$ at frequency $f_0$.

The power spectrum $P_r(f)$ of the TOA fluctuations is related to the the strain amplitude $h_c(f)$ by  \cite[][]{2009MNRAS.394.1945H}
\be
P_r(f)  = \frac{h_c^2(f)}{12 \pi^2 f^3}.
\ee

The rms of the residuals $\sigma_{{\rm GW},2}(T)$ over a time span $T$ is related to the power spectrum of the perturbations $P_r(f)$ by 
\be
\sigma_{{\rm GW},2}^2(T)= \int_{0}^{\infty} df H(f,T) P_r(f),  
\ee
where $H(f,T)$ is a high-pass filter that accounts for power that is removed by model fitting to the arrival times.

The rms of the residuals $\sigma_{{\rm GW},2}$ is most accurately determined by simulating the TOA perturbations associated with a GWB and then calculating the residual TOAs  and $\sigma_{{\rm GW},2}$.    For a gravitational wave background with $\alpha = -2/3$ 
\be
\label{eqn:gw_scaling}
\sigma_{{\rm GW},2}(T) \approx 1.3~{\rm ns}~A_{0,-15}  \left( \frac{T}{1~{\rm yr}}\right)^{5/3}, 
\ee
where $A_{0, -15} = A_0/10^{-15}$ is the characteristic strain at $f_0 =1$~yr$^{-1}$ and that the high pass filter is approximately unity for $f > 1/T$ and zero for $f < 1/T$. 
We note the scaling with $T$ is similar to that for a random walk in frequency (${\rm RW}_1$) for which $\sigma_{{\rm TN},2} \propto T^{3/2}$  is expected.  

The GWB was simulated using a large number of gravitational waves with wave amplitude, frequency, phase, polarization, and propagating direction randomly selected from appropriate PDFs.  In particular, the wave amplitude and frequency were selected from distributions consistent with Equation (\ref{eqn:ap_strain_amp}) using appropriate lower $f_\ell$ and upper frequency cut-offs.  Equation (\ref{eqn:gw_scaling}) is valid for all $f_\ell \ll 1/T$.  In simulations with $f_\ell = 1/(10\:T)$ we find that $\sigma_{{\rm GW},2}(T)  \approx 1/25~\sigma_{{\rm GW}}(T)$. 





\begin{deluxetable}{llrc}
\tablewidth{3in}
\tabletypesize{\footnotesize} \tablecolumns{4}
 \tablecaption{Previous Timing Campaigns\label{tab:time_sources}}
\tablehead{ \colhead{Reference} &\colhead{Objects} &  \colhead{$T_{\rm typ}$} & \colhead{Observation} \\
 \colhead{} &\colhead{} &  \colhead{(yr)} & \colhead{Type} }
  \startdata
  \multicolumn{4}{c}{Canonical Pulsars (CPs)}\\
  \hline
\cite{1980ApJ...237..206H} & 37  & 4  & TW\\ 
\cite{1985ApJS...59..343C}  & 27 &10  &TN\\ 
 \cite{1993MNRAS.261..883D} &45 & 4 & TW\\ 
\cite{1994ApJ...422..671A} &96  & 3 & TW \\
\cite{1994AJ....108..175F} & 1   &  6 & T${^*}$\\ 
\cite{1995MNRAS.277.1033D} & 45  & 7 & TN\\ 
\cite{2004MNRAS.353.1311H}  & 346, 27 MSP  & 20 & TSW \\ 
 \cite{2004MNRAS.354..811Z} & 2 & 1 &T \\
 \cite{2005MNRAS.363..929C} & 15, 1 MSP & 2 &T \\ 
\cite{2005ApJ...628L..45K} &  1  & 11 &T\\ 
\cite{2007ChJAA...7..521C} & 27  & 10 &S\\ 
\cite{2009MNRAS.400.1431M} & 6 RRAT& 6 &T\\ 
\cite{2010MNRAS.402.1027H} & 346, 30 MSP & 25 & TSW\\  
\hline
\multicolumn{4}{c}{Millisecond Pulsars (MSPs)}
\\ \hline 
\cite{1994ApJ...428..713K} & 2 & 2 & S\\ 
\cite{1997MNRAS.286..463B} & 4 & 3 &T\\ 
\cite{2002LommenThesis} & 4, 2 CP & 10 &S \\ 
\cite{2006MNRAS.369.1502H} & 15 & 2  &S\\
\cite{2006MNRAS.371..337O} & 1 & 4 &S\\ 
\cite{2007PhDT........14D} &  15 & 2&T\\   
\cite{2008ApJ...679..675V} &1 & 10 &T\\ 
\cite{2009PASA...26..103H}& 20 & 4 &T\\  
\cite{2009MNRAS.400..805L} & 1 & 14 &T\\ 
\cite{2009arXiv0906.4246V} &19& 10 &H\\ 
 \hline
 \multicolumn{4}{c}{Magnetars (MAGs)}
\\ \hline 
\cite{2000ApJ...535L..55W} & 1 &1 &S\\  
 \cite{2001ApJ...558..253K} &4 & 1 &S\\ 
 \cite{2002ApJ...567.1067G} & 5 & 1 &S\\ 
\cite{2002ApJ...564L..31G} & 1 & 2 &S\\ 
\cite{2007ApJ...663..497C} & 1 & 1 &S\\ 
\cite{2008AA...489..245D} & 3 & 1 &S\\ 
 \cite{2008ApJ...673.1044D} & 1 & 1 &S 
\enddata
\tablecomments{Timing campaigns used in this analysis. We list campaigns, class of objects studied, typical observing length $T_{\rm typ}$, and reported observable.   Horizontal lines divide campaigns that study predominantly canonical pulsars (CPs,  $1/6~{\rm s}^{-1} < \nu < 50$~s$^{-1}$), millisecond pulsars (MSPs; $\nu > 50$~s$^{-1}$), and magnetars (MAGs, $\nu < 1/6$~s). The rotating radio transients (RRATs) are rotating neutron stars that show spin properties similar to that of canonical pulsars.  The reported observation types are: TW, total rms residuals and whitened residuals; TN, timing noise, S,  higher order spindown terms (e.g., $\ddot{\nu}$), STW, spindown terms, total rms (after measurement of spin-down terms), and whitened residuals; H, harmonically related sinusoids and whitened residuals; T, only upper limits on levels of timing noise are reported; T$^{*}$, only the total rms is reported but timing noise is dominant contribution to rms residuals.}
\end{deluxetable}

\begin{deluxetable}{rrcc}
\tablewidth{3in}
\tabletypesize{\footnotesize} \tablecolumns{4}
 \tablecaption{Timing Noise in PSR~B1937$+$21\label{tab:tn_1937}}
\tablehead{ \colhead{$T$} & \colhead{$\sigma_{\rm RMS}$} & \colhead{$N_{\rm TOA}$} & \colhead{Ref.} \\
 \colhead{(yr)} &\colhead{($\mu$s)} &  \colhead{} & \colhead{}  }
\startdata
  1.0  &   0.15 & \nodata &    1  \\ 
  1.0  &   0.23 & 22   & 2 \\ 
  1.0  &  0.24  & 19 & 2 \\  
  1.0  & 0.32   & 16 & 2 \\
  1.0  & 0.24   & 16 & 2 \\ 
  1.0  & 0.21   & 18 & 2 \\
  1.0  & 0.21   & 14 & 2 \\ 
  1.0  & 0.19   & 23 &2 \\ 
1.2  &  0.21   & 13 & 2 \\ 
  1.5  & 0.17 & \nodata & 3 \\ 
2.0   & 0.25    & 47 & 2 \\  
    2.0  &  0.29   &  38 & 2\\ 
    2.0   & 0.20   & 38 & 2 \\ 
   2.2  & 0.20   &  38 & 2 \\  
2.3    & 0.19& \nodata & 4  \\  
2.4   & 0.20   &231&  5 \\  
2.7    & 0.32    & 85 & 6 \\ 
  4.0    & 0.20 &39 & 7 \\  
4.0    & 0.30  & \nodata & 8 \\  
4.0 &   0.41  &    85 & 2 \\
4.2 &   0.49    & 80 & 2\\
 4.4  &    0.27   & 168 & 9 \\ 
   8.2   &  0.94  & 440  & 2 \\ 
10.0    & 1.3~~   &  \nodata & 10 \\ 
16.8   &  9.3~~    &   387 & 10  \\ 
  20.0  &    112.0~ &400 & 11 \\  
23.3  &   24.2~   & 588 & 12  \\
24.0    &  27.4~    & \nodata  & 13
\enddata
\tablecomments{  Root mean square times of arrival for PSR B1937+21 for different observing programs.  
Column $T$ shows the observing span, column $\sigma_{\rm RMS}$ shows the total rms residuals, column $N_{\rm TOA}$ shows the number of times of arrival included in the analysis, and column Ref. shows the numbered references.
The references are:
(1) \cite{Man2008};  (2) \cite{1994ApJ...428..713K}; (3) \cite{Man2009}; (4) \cite{2007MNRAS.378..493Y}; (5) \cite{2006MNRAS.369.1502H}; (6) \cite{2007PhDT........14D}; (7) \cite{Dem2008};
(8) \cite{Thereau2008};
(9)  \cite{2009arXiv0906.4246V}; (10) \cite{2002LommenThesis}; (11) \cite{2004MNRAS.353.1311H}; (12) \cite{2009MNRAS.400..951V}; (13) \cite{Jan2008}.}
 \end{deluxetable}

\end{document}